%% file: From_Punishment_to_Protection_Charting_Six_Decades_of_Juvenile_Justice.tex
\documentclass[11pt]{article}
\usepackage{acl}
\usepackage{times}
\usepackage{latexsym}
\usepackage[T1]{fontenc}
\usepackage[utf8]{inputenc}
\usepackage{microtype}
\usepackage{graphicx}
\usepackage{booktabs}
\usepackage{amsmath}
\usepackage{enumitem}
\usepackage{longtable}
\usepackage{mdframed}
\usepackage{tabularx}
\usepackage{float}
\usepackage{afterpage}
\usepackage[section]{placeins}
\mdfdefinestyle{codeblock}{%
  backgroundcolor=gray!7,
  linecolor=gray!35,
  linewidth=0.5pt,
  innerleftmargin=8pt,
  innerrightmargin=8pt,
  innertopmargin=6pt,
  innerbottommargin=6pt,
  skipabove=4pt,
  skipbelow=4pt
}

\providecommand{\linenumbers}{}
\providecommand{\nolinenumbers}{}

\title{From Punishment to Protection: 
Charting Six Decades of U.S. Juvenile Justice Through Topic Modeling and LLM-Assisted Analysis}

\author{
  Nia E. George \\
  Sidwell Friends School \\
  \texttt{ngeorge27@sidwell.edu}
  \And
  Simeon Sayer \\
  Harvard University \\
  \texttt{ssayer@fas.harvard.edu}
}

\begin{document}
\maketitle

\begin{abstract}
Juvenile courts handle two very different kinds of
cases---young people accused of crimes, and children at
risk in their own families---and both streams have been
changing dramatically over the past fifty years.
This paper asks: what has shifted, and can computational
methods track that change at scale?

Topic modeling and LLM-assisted trend analysis is applied to 60{,}470
U.S.\ appellate opinions spanning 1970 to 2025,
identifying 182 distinct legal topics organized into
10 themes covering the full range of juvenile justice
litigation.
The results are striking.
Child welfare litigation tripled its share of the corpus.
Sex offender registration cases more than doubled.
Traditional punitive mechanisms---judicial transfer to
adult court, the juvenile death penalty---declined
sharply.
A new cluster of sentencing cases emerged after 2010,
reflecting landmark Supreme Court rulings that
fundamentally redrew the constitutional limits on
juvenile punishment.

Analysis also shows that legal vocabulary shifts decade by
decade: the language courts used in the 1970s can be
unrecognisable by the 2020s, even for the same underlying
legal question.
The fastest-growing area of the corpus has fractured into
dozens of jurisdiction-specific variants that no single
topic can capture.
In both cases, case counts alone would miss the full arc of
doctrinal change.

This paper demonstrates that large-scale, reproducible 
analysis of appellate case law---quantitative trends 
and doctrinal arcs alike---is possible and practically useful.
It also reveals critical risks that any AI-based decision
support tool used in juvenile justice and trained on
such corpus will encounter: temporal mismatch, vocabulary
drift, jurisdictional fragmentation, and the divergence
of delinquency and child welfare into two parallel legal
systems. Addressing these risks must be a fundamental 
requirement for any tool used in this domain.
\end{abstract}

\section{Introduction}

Juvenile courts sit at the intersection of child development,
public safety, and family regulation. Appellate opinions in
juvenile-related cases show how courts interpret youth
culpability, due process, sentencing limits, and
child-welfare obligations. Because these opinions span many
jurisdictions and decades, it is difficult to see long-run
patterns through traditional doctrinal reading alone.

Juvenile justice has also shifted substantially over time.
The Office of Juvenile Justice and Delinquency 
Prevention (OJJDP) data shows dramatic declines in 
youth arrests since the mid-1990s peak and sustained 
declines in delinquency caseloads through 2019 
\citep{puzzanchera2022}. Caseloads continued below 
pre-pandemic levels through 2023, though with year-to-year 
fluctuations following the COVID-19 period \citep{hockenberry2025}.
At the same time, litigation continues to shape practice
through major constitutional and statutory inflection
points, including the due-process cases of
\textit{Kent v.\ United States} \citep{kent1966},
\textit{In re Gault} \citep{gault1967}, and
\textit{In re Winship} \citep{winship1970}, and modern
limits on extreme youth punishment in
\textit{Roper v.\ Simmons} \citep{roper2005},
\textit{Graham v.\ Florida} \citep{graham2010},
\textit{Miller v.\ Alabama} \citep{miller2012}, and
\textit{Montgomery v.\ Louisiana} \citep{montgomery2016}
(see Appendix~\ref{app:inflections} for a full account
of these and the key federal statutes).

\paragraph{Why juvenile justice trends matter.}
Juvenile justice system decisions can shape a young person's
life at a critical stage---affecting whether they stay in
school, keep their family stable, and have opportunities
later on.
Developmental reform frameworks emphasize that accountability
should reflect adolescents' evolving capacities and promote
fairness and reduced re-offending \citep{nationalresearchcouncil2013}.
Evidence syntheses show that the effectiveness of intervention is
 heavily dependent on the quality of program design and implementation, 
making trend monitoring useful for directing resources
\citep{lipsey2009}.

\paragraph{Why multi-decade trend study is difficult.}
A multi-decade horizon covers periods of sharply
different approaches to juvenile justice---from the
due-process revolution to the punitive turn to recent
developmental reforms---but studying it systematically
is difficult \citep{fagan2008, schaefer2016, feld2017}.
Legal data availability is uneven: publication practices vary
by jurisdiction, older opinions are less consistently
digitized, and juvenile matters often involve confidentiality
norms that complicate systematic collection \citep{radice2018}.
Federal repositories such as PACER and state court
administrative offices have expanded online access to
historical opinions, but coverage remains uneven across
jurisdictions and time periods.
Even when texts are available, opinions are long,
citation-dense, and filled with boilerplate; terminology also
shifts as statutes and institutions change \citep{ariai2024}.

\paragraph{Why tracking legal trends matters for AI tools.}
Juvenile justice agencies increasingly rely on algorithmic
tools for prediction and triage, and recent advances in LLMs
have expanded interest in AI-assisted legal analysis.
When AI tools influence decisions about youth liberty,
over-reliance on model outputs is a documented risk,
and fairness and due process requires human oversight
\citep{dressel2018, berk2019, tolan2019}.
Tracking how courts reason about youth rights and system
obligations over time supports auditing and helps ensure that
AI tools reflect modern legal standards rather than outdated
punitive baselines \citep{tsarapatsanis2021}.

\paragraph{This study} addresses these challenges through topic modeling:
a computational method that reads a large collection of texts
and automatically identifies recurring clusters of words and
ideas. Each cluster---a \emph{topic}---represents a coherent
legal issue or doctrinal pattern. Related topics are grouped
into a \emph{theme}, providing a higher-level organizing
structure. Applied to legal opinions, topic modeling 
can surface the dominant legal questions in a corpus, track 
how their relative prominence shifts over time, and do so 
at a scale and consistency that manual doctrinal reading 
cannot match. The research questions below are organized 
around what such a pipeline can discover, how reliably it 
does so, and how its outputs can be made useful for legal 
scholarship.

\subsection*{Research Questions and Contributions}

\paragraph{Research questions:}
(RQ1)~What legal topics and themes characterize U.S.\
juvenile appellate case law across six decades?
(RQ2)~Which topics show the strongest growth or decline,
what drove those shifts and what does their underlying vocabulary
reveal about how juvenile justice doctrine has evolved?
(RQ3)~Can computational analysis of appellate case law produce findings that are reproducible, auditable, and reliable enough to support legal scholarship?
(RQ4)~What insights can be gained from the modeling process and 
analysis results that can serve as guidelines for  
AI-based decision support tools used in juvenile justice?

\paragraph{Contributions:}

\textit{Methodological.}
(1)~A \textbf{global--local BERTopic-based topic modeling architecture}
enabling consistent cross-decade topic comparison.
(2)~Built-in \textbf{topic-quality diagnostics} using UMass
coherence and nearest-neighbor lexical overlap.
(3)~A \textbf{two-stage LLM workflow} for  
theme discovery, topics organization and 
evidence-grounded trend analysis.

\textit{Research process.}
(4)~An \textbf{auditable output design and repeatable process} in which every
topic assignment, trend estimate, and narrative claim
is traceable to data at every pipeline stage.

\textit{Findings.}
(5)~\textbf{Empirical documentation} of six decades of
doctrinal rebalancing across 60,470 appellate opinions.
(6)~An \textbf{evidence-grounded AI risk framework},
identifying five guidelines for decision
support tools used in juvenile justice.

\section{Related Work}

\paragraph{Juvenile justice trends.}
Existing research measures juvenile justice trends
through three main approaches. Administrative data
tracks arrests, court caseloads, and placements.
OJJDP reporting documents steep declines in youth
arrests from the 1996 peak and sustained falls in
delinquency caseloads through the pre-pandemic period
\citep{puzzanchera2022, hockenberry2025}. Policy
scholarship traces the ideological cycles that drove
those trends---from the punitive expansion and transfer
legislation of the 1980s--1990s through developmental
reforms and constitutional constraints on youth
punishment \citep{fagan2008, feld2017, schaefer2016}.
Intervention research evaluates what programs and
practices reduce reoffending \citep{lipsey2009}.
Each of these approaches measures what the system
does or how it performs. None tracks how the legal
questions courts were asked to decide changed over
time. This study addresses that gap by analyzing
appellate doctrine directly, treating shifts in
the volume and content of court opinions as a
complementary measure of how juvenile justice
evolved across five decades.

\paragraph{Legal NLP and temporal topic modeling.}
Court opinions have been analyzed for argument structure,
outcome prediction, and citation patterns, with shared
test sets developed to measure legal language
understanding \citep{ariai2024, chalkidis2022}.
Topic modeling has the longest history of application to
legal corpora. \citet{livermore2017} applied LDA
\citep{blei2003} to the joint corpus of U.S.\ Supreme Court
and federal appellate opinions, measuring how the subject
matter of Supreme Court decisions differs from the broader
appellate corpus and tracking those differences over time---
the closest prior work to the present study in both method
and source material. LDA has also been applied to UK
legislation \citep{oneill2016}, European Court of Human
Rights decisions \citep{malakou2024}, and legislative
corpora in multiple languages \citep{ariai2024}.
Embedding-based approaches have more recently been
applied to legal texts: \citet{sohselective2024} compared
LSA with BERTopic \citep{grootendorst2022} on domain
name dispute decisions and European Court of Human Rights
cases, and LLM-assisted
topic classification of UK summary judgments has been
demonstrated using Claude \citep{vandenberghe2025}.
No prior work applies topic modeling to juvenile justice
case law specifically, and none addresses \emph{temporal
comparability across decades}: models fit separately per
period produce incomparable topic inventories. 

\section{Pipeline Overview}

\begin{figure}[t]
\centering
\includegraphics[width=\columnwidth]{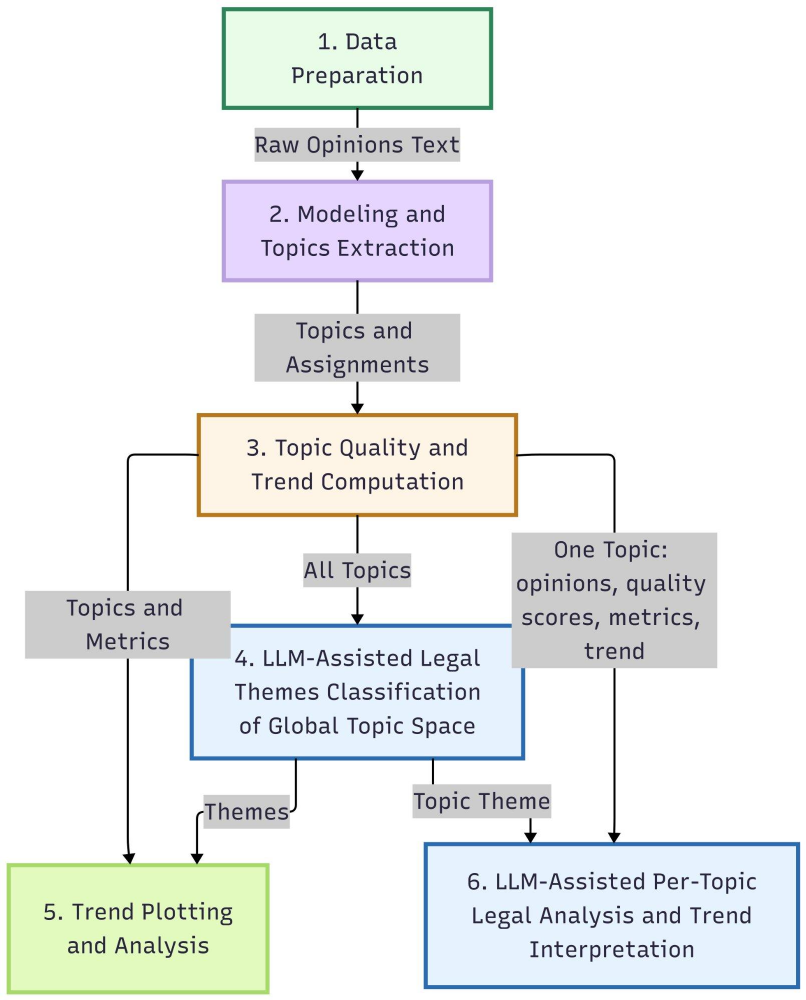}
\caption{End-to-end pipeline overview.}
\label{fig:pipeline}
\end{figure}

The analysis proceeds in six stages, as illustrated in
Figure~\ref{fig:pipeline}. First, juvenile
delinquency-related appellate opinions from 1970s to 2020s 
is collected
from CourtListener \citep{courtlistener}, a public U.S.\
judicial opinion repository, and cleaned through
text normalization, removal of legal boilerplate, and
detection of common legal phrases, producing a
decade-indexed corpus ready for topic modeling.
Second, a global BERTopic
model \citep{grootendorst2022} is fit to the full corpus to
establish a shared topic space, and separate per-decade models
are fit to capture period-specific structure; decade topics are
then aligned to the global inventory via top-word Jaccard
overlap. Third, topic quality is assessed using UMass coherence
and nearest-neighbor lexical overlap, and temporal trends are
computed as decade-normalized prevalence shares with linear
slopes to identify rising, stable, and declining topics. Fourth,
the full inventory of global topics is submitted to a
two-pass LLM workflow that discovers a set of legal themes 
from the topic inventory, and 
assigns every topic to exactly one theme. Fifth, theme- and
topic-level prevalence trajectories are analyzed across the
1970s through the 2020s to identify the strongest movers in
each direction. Sixth, structured evidence bundles, consisting of
original opinions texts and quality and trend metrics 
are assembled for selected topics of interest and 
submitted to the LLM for per-topic legal analysis. 
Each stage of this pipeline is 
described in detail in the sections that follow.

\section{Data}
Juvenile courts handle two very different kinds of
cases. The first is the delinquency case: a young
person is accused of an offense and the court
decides whether they are culpable and what should
happen next---with an emphasis on rehabilitation
rather than punishment. The second is the child welfare
case: the state steps in because a child may be
unsafe at home, and the court oversees decisions
about removal, family services, and placement. Both
types of cases generate appellate opinions. 
The corpus captures both streams.
A single opinion may address how a delinquency
hearing was conducted, whether a youth should be
tried as an adult, what sentence is constitutionally
permitted, or whether a child welfare agency followed
the rules before removing a child from their family.
All of these are treated as juvenile justice case law
throughout the analysis.

\subsection{Corpus Construction}
Appellate opinions are collected from CourtListener using
their public REST APIs. To ensure balanced temporal coverage 
and reduce retrieval bias from API paging, queries were 
executed in multiple time slices using decade blocks and 
within-decade multi-year windows, preventing any single high-volume 
period from dominating the sample. After de-duplication 
and filtering, \textbf{60{,}470 opinions}
spanning 1970--2025 formed the corpus (Table~\ref{tab:corpus}).

\begin{center}
\small
\captionof{table}{Corpus distribution by decade.}
\label{tab:corpus}
\begin{tabular}{lrr}
\toprule
\textbf{Decade} & \textbf{Opinions} & \textbf{\%} \\
\midrule
1970s &  7,030 & 11.6 \\
1980s & 10,330 & 17.1 \\
1990s & 11,024 & 18.2 \\
2000s & 11,057 & 18.3 \\
2010s & 11,000 & 18.2 \\
2020s & 10,029 & 16.6 \\
\midrule
\textbf{Total} & \textbf{60,470} & \textbf{100} \\
\bottomrule
\end{tabular}
\end{center}

\subsection{Preprocessing}

Opinion text was cleaned by removing punctuation-only
tokens, citation fragments, person names (identified
using spaCy \citep{honnibal2020}), and formatting
artifacts from heterogeneous sources (HTML conversion,
PDF extraction, scanning). Common multi-word legal
phrases were joined into single tokens using a phrase
model trained once on the full corpus and applied
uniformly to every decade slice \citep{rehurek2010},
so that expressions like \textit{termination of
parental rights} are treated as one concept throughout.
A conservative legal stop-term list then suppressed
high-frequency boilerplate while preserving doctrinal
vocabulary. The same configuration was applied
identically to all six decade slices so that trend
differences reflect genuine doctrinal change, not
preprocessing variation.

\subsection{Data Quality}

All 60,470 opinions were retained after preprocessing,
with no documents lost to filtering. Roughly 65\% of
raw word tokens survived cleaning, and retention rates
were consistent across all six decades
(62.8\%--67.3\%), which means that differences in
topic trends are unlikely to be caused by the cleaning
process treating older or newer opinions differently.
A small number of text-extraction errors were left in
place rather than removed, since aggressive filtering
risked deleting genuine legal content; at the scale of
60,470 opinions, isolated noise has little effect on
topic discovery.

\section{Modeling and Topic Extraction}
\label{sec:modeling}

Topic modeling required a method that could represent
substantive legal issues in long appellate opinions,
stay consistent across five decades of changing
vocabulary, and work within the multi-stage pipeline.
Both word-count-based models (LDA, Top2Vec/Doc2Vec)
and embedding-based models (BERTopic, Top2Vec/MiniLM)
were tested on a 6,000-opinion pilot subset sampled
to be roughly balanced across decades, and assessed on
three criteria: how coherent and interpretable the topics
were, how distinct the topics were from one another,
and how well each model fit the pipeline's requirements.
Full comparison results are in
Appendix~\ref{app:modelcomp}.

\subsection{Model Selection}

BERTopic using MPNet sentence embeddings,
UMAP \citep{mcinnes2018umap}, and HDBSCAN
\citep{mcinnes2017hdbscan} was selected based on superior
performance across all three criteria. Over the
1970s--2020s decade slices, BERTopic/MPNet achieved a
mean c$_v$ coherence of 0.493 (SD 0.013; range
0.474--0.504) \citep{roder2015}, outperforming LDA
(0.382 $\pm$ 0.004), Top2Vec/MiniLM (0.467 $\pm$ 0.035),
and a LegalBERT variant (0.441 $\pm$ 0.058). Topic
distinctiveness was also stronger: mean nearest-neighbor
cosine similarity was approximately 0.40 for BERTopic/MPNet
versus 0.66 for Top2Vec/MiniLM, indicating lower topic
redundancy. Cross-decade continuity, measured by best-match
cosine similarity between topic word signatures in adjacent
decades, was highest for BERTopic/MPNet (mean $\approx$
0.51), exceeding both Top2Vec variants. Operationally,
BERTopic supported hierarchical topic reduction, reliable
export of document-topic assignments and c-TF-IDF
signatures, and straightforward assembly of evidence
bundles for the LLM-assisted stages. 

\subsection{Global--Local Architecture}
\label{sec:globallocal}

To support both cross-decade comparability and
decade-local specificity, a two-level design is used.
A single global model is fit over the full 60,470-opinion
corpus to establish a shared topic space, producing 182
substantive topics after hierarchical reduction. Separate
per-decade models are then fit independently for each
of the six decade slices, producing 235--333 topics per
decade after reduction. To link decade-local topics to
the global inventory, the pipeline computes the Jaccard
similarity between each local topic's top-word set and
each global topic's top-word set. A local topic is
\textit{mapped} to the global topic with the highest
Jaccard score if that score is at least 0.20; topics
that fall below this threshold are \textit{unmapped}.
This threshold is intentionally moderate: in a legal
corpus where different courts across different time periods 
may use different terminology
for the same doctrine, exact word overlap is rare, and
0.20 represents a meaningful minimum of shared
vocabulary. The result is consistent cross-decade
prevalence estimation from mapped topics, and
isolation of unmapped topics that reflect
era-local legal developments.


\section{Topic Quality and Trend Computation}

A topic model produces clusters automatically, but not
all clusters are equally meaningful. In a legal corpus,
some topics may consist of boilerplate language that
appears across many case types rather than a coherent
legal issue; others may be near-duplicates of each other
because courts in different jurisdictions use slightly
different words for the same doctrine. Without a way to
screen for these problems before interpretation begins,
the trend analysis risks treating noise as signal.
Separately, raw document counts cannot be compared
across decades because each decade slice is a different
size---a topic that appears in 500 opinions out of 7,000
in the 1970s is more prominent than the same count out
of 11,000 in the 2000s. Both problems require a
quantitative assessment of topics' quality to guide their 
use in further analysis.
The subsections below describe how the pipeline produces
the topic inventory, assesses quality, and computes
trends in a way that is transparent and auditable at
every step.

\subsection{Global Model: Topic Inventory}

\begin{figure}[t]
\centering
\includegraphics[width=\columnwidth]{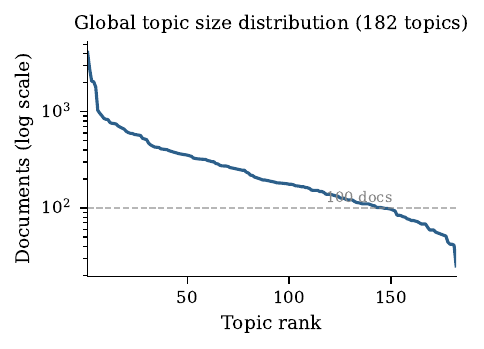}
\caption{Ranked topic size distribution for the 182
global topics (log scale). The steep initial drop
followed by a long flat tail is characteristic of
legal corpora.}
\label{fig:topicsizes}
\end{figure}

The global BERTopic model was fit on the full
60,470-document corpus and reduced to 182 substantive
topics. These 182 topics collectively cover 33,077 corpus
opinions (54.7\%); the remaining 45.3\% were not assigned
to any substantive topic and are excluded from prevalence
estimates. Figure~\ref{fig:topicsizes} shows the ranked
topic size distribution on a log scale. The curve
descends steeply from T0 (4,109 documents) to the
10th-ranked topic, then flattens gradually through the
long tail, where most topics average around 60 documents.
This long-tailed distribution is typical of large legal
corpora: the model neither collapses opinions into a
handful of oversized clusters nor fragments them into
hundreds of near-empty topics.

\subsection{Decade-Local Models and Cross-Decade Alignment}

\begin{figure}[t]
\centering
\includegraphics[width=\columnwidth]{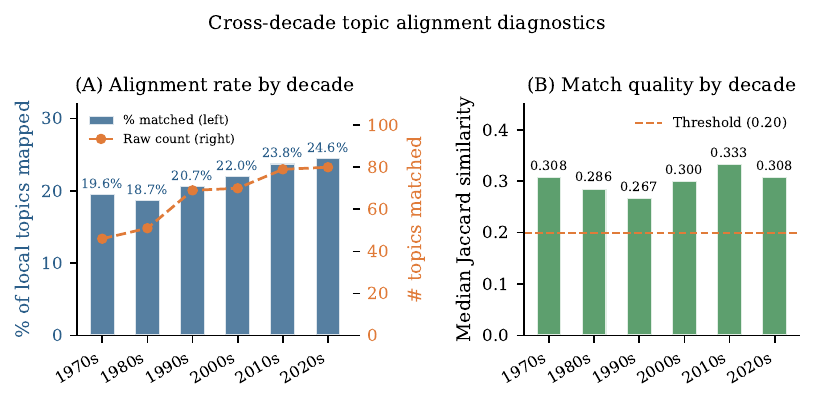}
\caption{Cross-decade topic alignment. (A)~Match
rate and count by decade. (B)~Median Jaccard
similarity of accepted matches; dashed line = 0.20
threshold.}
\label{fig:alignment}
\end{figure}

Six decade-local BERTopic models were fit independently,
producing 235--333 topics per decade after reduction.
Figure~\ref{fig:alignment} reports the median Jaccard
similarity of accepted matches (decade topics that 
mapped to a global topic) per decade (Panel B). 
The values range from 0.267 (1990s) to 0.333 (2010s) and remain
stable across all six decades, indicating that the
quality of accepted matches---not just their count---is
consistent over time. Alignment rates (Panel A) improve from
19.6\% in the 1970s to 24.6\% in the 2020s, reflecting
progressive standardization of legal vocabulary as
constitutional and statutory doctrine consolidated.
Together, these diagnostics confirm a durable common core
of legal topics that recurs in every decade.

\subsection{Topic Quality Diagnostics}

\begin{figure}[t]
\centering
\includegraphics[width=\columnwidth]{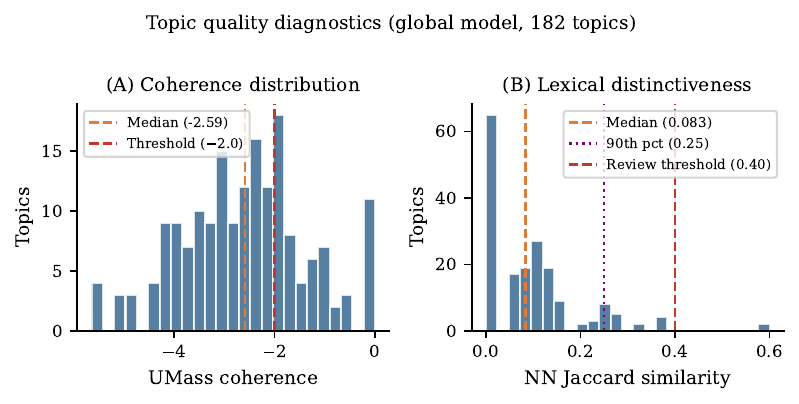}
\caption{Topic quality diagnostics for 182 global
topics. (A)~UMass coherence; dashed line = $-2.0$
usability threshold. (B)~Nearest-neighbor Jaccard
similarity; dashed line = 0.40 review threshold.}
\label{fig:quality}
\end{figure}

Figure~\ref{fig:quality} shows the distribution of both
quality metrics for the 182 global topics.

\textbf{UMass coherence} (Panel A) measures whether a
topic's top words tend to co-occur in the same documents.
Scores are typically negative; values at or above
$-2.0$ are comparatively strong for a sparse legal
corpus, values in the $-2.5$ to $-3.0$ range are often
usable though noisier, and values below $-4.0$ typically
indicate weak or artifact-prone topics
\citep{mimno2011}. The global model achieves a median
coherence of $-2.59$, with 55 topics (30.4\%) at or
above the $-2.0$ usability threshold. The distribution
spans from below $-5.5$ to $0$, reflecting the range of
topical clarity in a corpus where procedural boilerplate
is pervasive alongside substantive doctrinal language.

\textbf{Nearest-neighbor Jaccard similarity} (Panel B)
measures how much each topic's top-10 words overlap
with its most similar neighbor. Lower values indicate
more distinct topics. Values below 0.15 suggest clear
lexical separation; values in the 0.15--0.30 range
reflect moderate overlap common among related legal
themes; values above 0.40 warrant inspection for
redundancy. The distribution is strongly right-skewed:
the median is 0.083 and the 90th percentile is 0.25,
indicating that the vast majority of topics are
lexically well-separated. Only 2 of 182 topics (1.1\%)
reach the 0.40 review threshold, confirming that
topic redundancy is minimal.

These diagnostics are used as interpretive guides, not
hard cutoffs. Because appellate opinions extensively
reuse statutory language and offense labels, coherence
scores in the $-2.5$ to $-3.0$ range can still
correspond to legally interpretable topics. All 182
topics are retained in the prevalence table; for
LLM-assisted thematic interpretation and evidence 
packaging, topics are prioritized using a 
combination of coherence,
substantive legal relevance, and size.

\begin{figure}[b]
\centering
\includegraphics[width=\columnwidth]{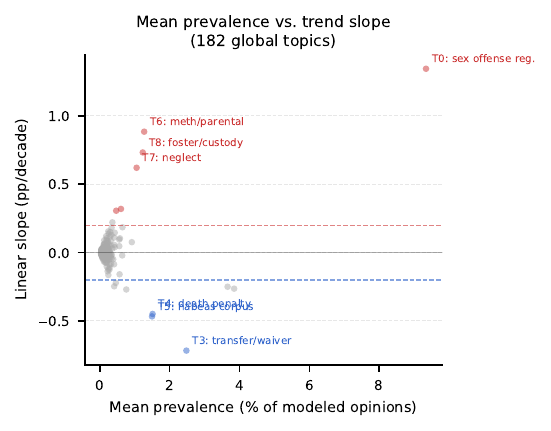}
\caption{Mean prevalence vs.\ linear slope for all
182 global topics. Red = rising; blue = falling;
gray = stable.}
\label{fig:scatter}
\end{figure}

\subsection{Temporal Trend Computation}

For each global topic and each decade, the count of
assigned opinions is divided by the total assigned
opinions in that decade, producing a directly comparable
decade-normalized share. A linear slope is then fit for
each topic through its six decade-normalized shares,
expressed in percentage points per decade.

Figure~\ref{fig:scatter} plots mean prevalence against
linear slope for all 182 topics. The dense gray cluster
near the origin reflects the stable majority: 167 of
182 topics (91.5\%) have an absolute slope below 0.20
pp/decade. The seven rising topics (red) and eight
falling topics (blue) spread outward from this cluster,
with T0 (sex offender registration) isolated in the
upper right at a mean prevalence of 9.3\% and a slope
of $+1.34$ pp/decade---the largest topic and the
strongest riser in the corpus. These labeled outliers
identify the topics driving the trend analysis in
Section~\ref{sec:themeinventory}.


\section{LLM-Assisted Theme Classification}
\label{sec:themeclassification}

With 182 global topics produced by the BERTopic model,
direct interpretation of the full topic inventory would
be unwieldy. This next stage organizes the inventory into a
compact, human-interpretable legal taxonomy using a
large language model. The classification is based
entirely on topic metadata---no opinion texts are
included. This keeps the prompt short enough to work
well and avoids document-level noise at a structural
classification step.

\subsection{Two-Pass Classification}

The classification runs in two passes to produce a
stable taxonomy without sacrificing coverage.

\textbf{Pass 1: Theme induction.}
A representative set of topics is constructed from
higher-quality topics (by UMass coherence), the largest
topics by document count, and topics from detected
juvenile-specific legal areas. Each topic is submitted
to the model as a short record containing: its top 12
c-TF-IDF words, total document count, coherence score,
nearest-neighbor overlap score, trend direction
(rising, stable, or falling), and how many decades it
appears in. The LLM uses this set to propose a
legal theme taxonomy of 6--14 themes, broad enough to
cover the corpus but legally meaningful. The LLM is
prompted as a legal taxonomy expert in U.S.\ appellate
law and juvenile justice and instructed to return
structured JSON only.

\textbf{Pass 2: Topic assignment.}
In a second pass, the full inventory of 182 topics 
is submitted along with the
fixed Pass 1 taxonomy. The model assigns each topic to
exactly one existing theme and returns a confidence
score (0--1) and a one-sentence rationale per
assignment. Pass 2 also produces theme-level metrics,
top-five topic lists by document count, and a short
legal summary for each theme. 


\section{LLM-Assisted Per-Topic Legal Analysis}
\label{sec:pertopicanalysis}

The second LLM inference stage applies evidence-grounded
analysis to individual topics rather than to the full
topic inventory. For each selected topic, a structured
evidence bundle is assembled and submitted to the LLM.
Outputs are treated as informed hypotheses grounded in
the supplied opinions, not as authoritative legal
conclusions.

\subsection{Evidence Packaging}

The evidence bundle has three components. First,
\textit{topic metadata}: the topic identifier, top-ranked
c-TF-IDF terms with scores, decade-by-decade
document counts and normalized corpus-share values.
Second, a \textit{stratified opinion sample} drawn from
three pools. The first pool is the global topic
itself---opinions assigned to the target global
topic across the full corpus. The second pool is
the decade-local mapped topics: local topics from
each decade slice whose Jaccard similarity to the
global topic met the 0.20 threshold. The third pool 
is the sub-theme topics: unmapped local 
topics that belong to the same legal territory as 
a global topic but fall below the alignment threshold. 
Each sub-theme is defined by a set of seed global topics. A decade-local unmapped topic qualifies by 
meeting two conditions. First, its best Jaccard 
match must be one of the seed topics. Second, its 
top-words list must either contain sub-theme 
vocabulary or show residual Jaccard overlap above 
a minimum floor. This third
pool captures era-specific legal vocabulary that
the global topic does not represent. Third,
\textit{full opinion texts} for all selected
opinions, without truncation, giving the model the
same source material a human legal reviewer would
consult.

\subsection{Analysis Mode and Prompt Architecture}

The prompt instructs the model to produce a
\textit{trend mode} analysis: tracing how the
relevant doctrine changed across decades by
identifying the dominant legal questions in each
era, explaining how doctrine shifted, attributing
the trend to specific legal and institutional
events, and noting which cases were most
consequential. The model is explicitly instructed
to base every claim on the supplied opinion texts,
not on prior legal knowledge.

The prompt uses a two-layer structure. The system
prompt establishes the model's role as a legal analyst
and instructs it to ground all claims in the supplied
opinion texts. The user prompt delivers the evidence
package and prescribes six required output sections
by name---Topic Overview, Decade-by-Decade Analysis,
Cross-Decade Changes, Drivers of the Trend, Key Cases
and Their Role, and Boundary Cases and Evidence
Limits. This fixed structure prevents the model from
substituting general legal knowledge for documented
judicial reasoning: every claim must be tied to the
supplied opinions rather than to what the model
already knows about the law. Full outputs are written
to structured files for audit; the complete outputs
for the three topics analyzed in this study are
reproduced in Appendix~\ref{app:pertopicoutputs}.

\section{Results}
\label{sec:results}

\begin{figure*}[t]
\centering
\includegraphics[width=\textwidth]{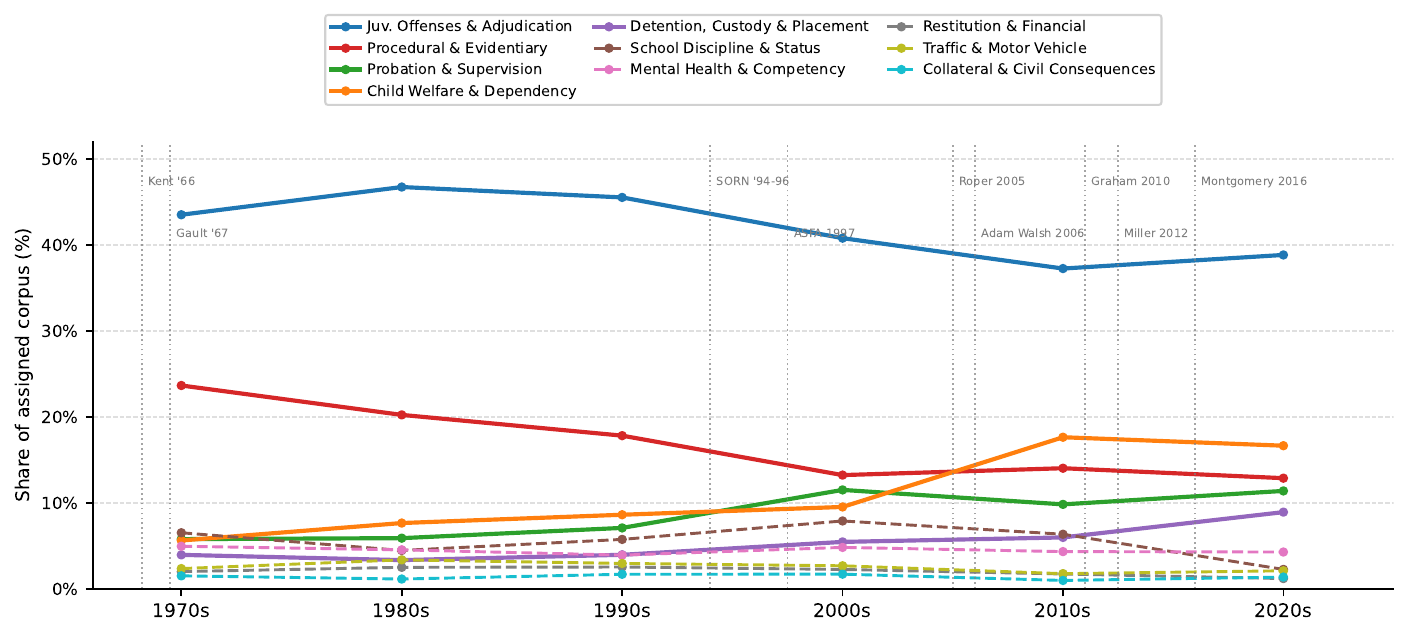}
\caption{All 10 legal theme shares (\% of modeled opinions)
by decade. Solid lines: high-volume themes; dashed lines:
low-volume themes. Policy markers: \textit{Kent} (1966),
\textit{Gault} (1967), SORN (1994--96), ASFA (1997),
\textit{Roper} (2005), Adam Walsh (2006),
\textit{Graham} (2010), \textit{Miller} (2012), and
\textit{Montgomery} (2016). \textit{Kent} and \textit{Gault}
predate the study window but shaped the 1970s corpus.
See Appendix~\ref{app:inflections} for details on all
policy markers.}
\label{fig:themes}
\end{figure*}

The results are organized in four layers, each drilling
deeper into the data. Section~\ref{sec:themeinventory}
describes the 10 high-level \textit{themes}---each a
named grouping of many related topics---and their
fifty-year prevalence trends: the widest lens, showing
where the system's appellate attention shifted overall.
Section~\ref{sec:topictrends} moves to the
\textit{global topic} level, examining the 12 largest
individual topics with their representative
vocabulary, document counts, and trend slopes.
Section~\ref{sec:decadelocal} zooms into
\textit{decade-local} topics to show how the same
legal doctrine was litigated differently in each era---capturing
vocabulary that the global model cannot represent,
including jurisdiction-specific statute language and
concepts that only emerged in recent decades.
Section~\ref{sec:pertopic} demonstrates how the LLM-assisted
per-topic analysis combines all three layers---global
trend data, decade-local specificity, and full opinion
texts---to produce an evidence-grounded legal narrative
for a single topic, illustrating what the pipeline can
recover that the statistical models alone cannot.

\subsection{Theme Inventory and Their Trends}
\label{sec:themeinventory}
Ten legal themes spanning all 182 topics were identified by 
the LLM assisted classification step (Table~\ref{tab:themes}). 

\begin{table}[t]
\centering
\small
\caption{10-theme taxonomy: coverage and trend.
Coverage = share of 33,077 modeled opinions.}
\label{tab:themes}
\begin{tabular}{lrr}
\toprule
\textbf{Theme} & \textbf{Cov.} & \textbf{Trend} \\
\midrule
Juv.\ Off.\ \& Adjudication      & 46.5\% & Stable  \\
Procedural \& Evidentiary         & 15.3\% & Falling \\
Probation \& Supervision          &  9.6\% & Rising  \\
Child Welfare \& Dependency       &  9.1\% & Rising  \\
Detention, Custody \& Placement   &  5.5\% & Rising  \\
School Discipline \& Status       &  4.5\% & Stable  \\
Mental Health \& Competency       &  3.2\% & Stable  \\
Restitution \& Financial          &  2.6\% & Stable  \\
Traffic \& Motor Vehicle          &  2.2\% & Stable  \\
Collateral \& Civil Consequences  &  1.5\% & Stable  \\
\bottomrule
\end{tabular}
\end{table}

Figure~\ref{fig:themes} shows how each theme's share
changed across six decades. The overall shift is from
cases about how youth enter the system---adjudication,
procedures, school conduct---toward cases about what
happens afterward: supervision, placement, and child
welfare. Three themes grew, one fell steadily, and the
rest remained roughly stable.

\begin{figure*}[t]
\centering
\includegraphics[width=\textwidth]{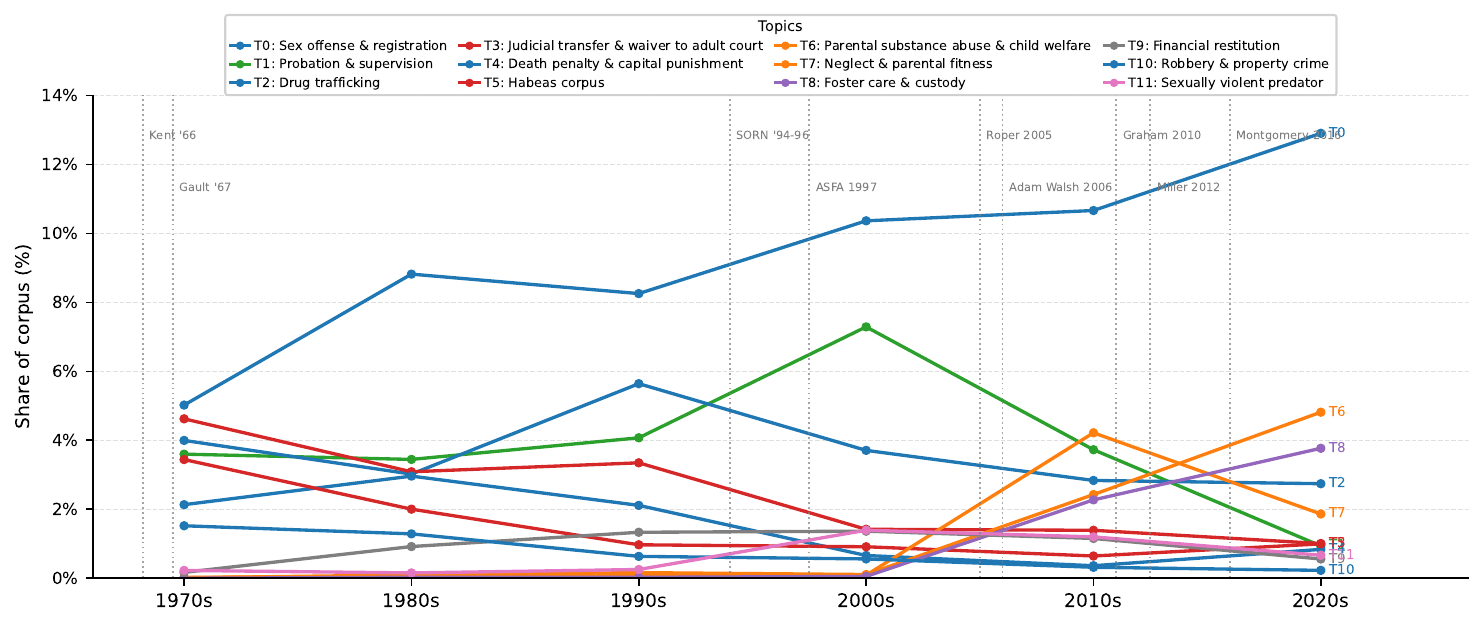}
\caption{Decade-normalized prevalence trajectories for
the 12 largest global topics. Line colors match the theme
color key in the upper legend. Topic IDs are annotated at
the 2020s endpoint. Policy markers: \textit{Kent} (1966),
\textit{Gault} (1967), SORN (1994--96), ASFA (1997),
\textit{Roper} (2005), Adam Walsh (2006),
\textit{Graham} (2010), \textit{Miller} (2012), and
\textit{Montgomery} (2016).
The 2020s endpoint covers 2020--2025 only and is
treated as preliminary.}
\label{fig:top12}
\end{figure*}

\paragraph{Child Welfare and Dependency.}
The fastest-growing theme, rising from 5.6\% in the
1970s to 16.7\% in the 2020s. Two developments drove
this: ASFA \citep{asfa1997} imposed strict timelines
on family reunification, producing a wave of appeals
over whether agencies followed the rules (the theme
jumps from 2.3\% to 9.5\% after the ASFA marker in
Figure~\ref{fig:themes}); and the methamphetamine
epidemic generated a surge in child removals due to
parental drug use, with court cases over removal
decisions and reunification plans
\citep{nationalresearchcouncil2013}.

\paragraph{Juvenile Offenses and Adjudication.}
The largest theme throughout, holding around 22--27\%.
Its share eased from a peak in the 1990s to 23.9\% by
the 2020s as appellate attention shifted to supervision
and welfare. Supreme Court rulings prohibiting the
juvenile death penalty (\textit{Roper}) and limiting
life-without-parole (\textit{Graham}, \textit{Miller},
\textit{Montgomery}) also redirected some appeals toward
resentencing hearings.

\paragraph{Probation, Supervision, and Revocation.}
Rose from around 4\% to 7.9\% in the 2000s as
supervision programs expanded, then fell back to
5--6\% as juvenile arrest rates declined and diversion
programs handled more cases outside formal court
\citep{puzzanchera2022, hockenberry2025}.

\paragraph{Detention, Custody, and Placement.}
Below 2\% for most of the study period, then jumped
to 6.6\% in the 2020s, reflecting more disputes over
where youth are placed and under what conditions,
paralleling the expansion of child welfare litigation
\citep{schall1984, nationalresearchcouncil2013}.

\paragraph{School Discipline, Truancy, and Status Offenses.}
Peaked at 3.9\% in the 2000s and fell to 0.9\% by
the 2020s, as states moved away from bringing
school behavior and status offenses into formal court
proceedings \citep{hockenberry2025, jjdpa1974}.

\paragraph{Procedural and Evidentiary Issues.}
This theme, which covers procedural safeguards, 
evidentiary rules, expungement, contempt, and 
other process-focused matters in juvenile proceedings 
shows the steepest sustained decline: from 12.6\% in the
1970s to 6.7\% in the 2020s. Once foundational
due-process rules were settled by \textit{Kent},
\textit{Gault}, and \textit{Winship}, those
questions generated fewer new appeals. 

\paragraph{}The four remaining themes---
Mental Health, Restitution,
Traffic, and Collateral Consequences---each stayed
below 4\% and stable; see Appendix~\ref{app:taxonomy}
for full details.


\subsection{Top Topics and Their Trends}
\label{sec:topictrends}

Table~\ref{tab:toptopics} lists the 12 largest global
topics by document count with their representative
terms, document count, theme assignment, linear slope,
and trend direction. Figure~\ref{fig:top12} plots their
decade-normalized prevalence trajectories. Topic lines
are colored by theme, matching the theme color key in
Figure~\ref{fig:themes}.

\begin{table*}[t]
\small
\centering
\caption{Top 12 global topics by document count.
Slope = linear change in decade-normalized share in
percentage points per decade step.
Trend: Rising $\geq +0.20$, Falling $\leq -0.20$,
otherwise Stable.}
\label{tab:toptopics}
\setlength{\tabcolsep}{4pt}
\begin{tabular}{clrp{3.2cm}rr}
\toprule
\textbf{ID} &
\textbf{Top terms} &
\textbf{Docs} &
\textbf{Theme} &
\textbf{Slope} &
\textbf{Trend} \\
\midrule
T0  & sex offender, rape, sexual assault
    & 5{,}787 & Juvenile Offenses         & $+1.34$ & Rising  \\
T1  & probation, supervision, revoke
    & 2{,}368 & Probation \& Supervision  & $-0.26$ & Falling \\
T2  & cocaine, heroin, narcotic
    & 2{,}212 & Juvenile Offenses         & $-0.25$ & Falling \\
T3  & transfer, jurisdiction, waiver
    & 1{,}424 & Procedural \& Evidentiary & $-0.72$ & Falling \\
T4  & death penalty, capital punishment
    &   886   & Juvenile Offenses         & $-0.45$ & Falling \\
T5  & habeas corpus, extradition, parolee
    &   828   & Procedural \& Evidentiary & $-0.47$ & Falling \\
T6  & methamphetamine, sobriety, parental right
    &   791   & Child Welfare             & $+0.88$ & Rising  \\
T7  & neglect, adjudicator, parental fitness
    &   676   & Child Welfare             & $+0.62$ & Rising  \\
T8  & custody, caseworker, foster care
    &   640   & Detention \& Custody      & $+0.73$ & Rising  \\
T9  & restitution, economic loss, victim
    &   588   & Restitution               & $+0.08$ & Stable  \\
T10 & robbery, bank, armed
    &   430   & Juvenile Offenses         & $-0.27$ & Falling \\
T11 & sexually violent predator, commitment
    &   413   & Mental Health             & $+0.19$ & Stable  \\
\bottomrule
\end{tabular}
\end{table*}

The trajectories reveal striking divergence across the
top 12 topics. T0 (sex offense registration) is both
the largest topic and the strongest riser, climbing
from 5.0\% in the 1970s to 12.9\% in the 2020s
($+1.34$ pp/decade). A modest rise in the 1980s
predates registration law: it reflects the expansion
of child sexual abuse prosecution following amendments
to the Child Abuse Prevention and Treatment Act
\citep{capta1984}, which drove state-level legislative
change in evidentiary standards and mandatory reporting
before any registration infrastructure existed. The
sharp acceleration from the 1990s onward tracks the
expansion of sex offender registration and notification
(SORN) law: a wave of federal mandates that required
every state to build and publish registries of convicted
sex offenders. The Jacob
Wetterling Act \citep{wetterling1994}, Megan's Law
\citep{meganlaw1996}, and the Adam Walsh Act (AWA)
\citep{adamwalsh2006} each generated new rounds of
appellate litigation over retroactive application,
lifetime registration for juvenile offenders, and
residency restrictions.
See Appendix~\ref{app:inflections} for a full account
of these statutory inflection points.

T6 (parental substance abuse and child welfare) is the
second-strongest riser ($+0.88$ pp/decade), with a
trajectory that is especially striking: near-zero
prevalence in the 1970s and 1980s, then a sharp jump
in the 2010s to 2.4\% and 4.8\% in the 2020s. This
pattern maps directly onto the methamphetamine epidemic
and the ASFA mandate \citep{asfa1997}, which together
generated a large wave of termination-of-parental-rights
proceedings. T7 (neglect and parental fitness, $+0.62$)
and T8 (foster care and custody, $+0.73$) follow
similar trajectories, capturing the procedural and
substantive sides of the same child welfare expansion.

On the falling side, T3 (judicial transfer and
waiver, $-0.72$ pp/decade) has the steepest downward
slope in the corpus, falling from 4.6\% in the 1970s
to 1.0\% in the 2020s. Its decline reflects the
replacement of discretionary judicial waiver hearings
by automatic transfer statutes across most states
from the late 1980s onward \citep{feld2017}: as
automatic transfer removed the need for individualized
hearings, the rich appellate jurisprudence around
waiver standards became correspondingly less prominent.

T4 (death penalty and capital punishment, $-0.45$ pp/decade) shows a trajectory that reflects the full arc of the punitive era rather than a single ruling. The topic peaks in the 1980s at 2.96\% as juvenile death penalty cases expanded alongside the broader punitive turn, then falls sharply through the 1990s and 2000s as constitutional scrutiny mounted and state-level abolitions accumulated before any federal ruling. \textit{Roper v.\ Simmons} in 2005 formalized what the data already show as an established trend, rather than triggering it. The modest uptick in the 2020s (0.36\% to 0.84\%) reflects resentencing hearings generated by \textit{Miller} and \textit{Montgomery}, which required courts to revisit sentences imposed under standards that no longer apply.

T1 (probation, supervision, and revocation, $-0.26$ pp/decade) shows a pronounced non-monotonic arc that tracks the expansion and subsequent contraction of formal juvenile supervision. The topic holds steady through the 1970s and 1980s before rising to a sharp peak of 7.3\% in the 2000s, reflecting the proliferation of intensive supervision programs, electronic monitoring requirements, and probation revocation proceedings that accompanied the punitive expansion \citep{feld2017}. From the 2010s onward the trend reverses sharply, falling to 3.7\% and then to 0.94\% in the 2020s, consistent with documented declines in juvenile arrest rates and the sustained shift toward diversion and informal processing that kept more cases outside formal court proceedings \citep{puzzanchera2022, hockenberry2025}. The linear slope understates the magnitude of this reversal: the 2000s peak is nearly eight times the 2020s level.

T2 (drug trafficking, $-0.25$) and T10 (robbery,
$-0.27$) declined in parallel with falling juvenile
arrest rates for drug and property crimes over this
period \citep{puzzanchera2022}. T5 (habeas
corpus, $-0.47$) declined in parallel with the general
contraction of traditional habeas petitions. T9 (restitution) and
T11 (sexually violent predator civil commitment)
remained stable, representing persistent but bounded
doctrinal areas that did not grow or shrink materially
over the study period.


\subsection{Decade-Local Structure}
\label{sec:decadelocal}

The global trends in Sections~\ref{sec:themeinventory}
and~\ref{sec:topictrends} show how the overall shape
of juvenile justice litigation changed across fifty
years. This section looks inside each decade to ask
what courts were actually arguing about in a given era,
and whether that matches what the global model sees.
Across most legal areas the answer is yes: courts
converged on shared vocabulary as doctrine stabilised
\citep{feld2017}. The exception is sex offense
registration, which generates persistently era-specific sub-theme
topics in every decade because the law kept
changing in ways that produced new, state-specific
vocabulary each time.

The global T0 label captures the total volume of
this litigation. The decade-by-decade causes and
doctrinal detail behind the sub-theme topic pattern
are examined in the per-topic LLM analysis of T0
in Section~\ref{sec:t0analysis}. The implications
for AI tools are discussed in
Section~\ref{sec:aiwarning}.


\subsection{LLM-Assisted Per-Topic Legal Analysis}
\label{sec:pertopic}

The three preceding sections established what the
corpus contains at increasing levels of resolution:
themes, global topics, and decade-local clusters.
This section shows what the LLM layer adds when all
three are combined into a single evidence bundle and
supplied to the model in trend mode---tracing how
doctrine evolved decade by decade rather than just
measuring its volume. Three topics are analyzed: T0
(sex offense adjudication and registration), the
largest and fastest-rising topic; T2 (juvenile drug
trafficking), a falling topic driven by successive
drug crises; and T6 (parental substance abuse and 
child welfare and ), the second-strongest riser. Each
analysis is accompanied by a timeline summarizing the
doctrinal arc. Full outputs are in
Appendix~\ref{app:pertopicoutputs}.

\subsubsection{Topic T0: Juvenile Sex Offense
Adjudication and Registration}
\label{sec:t0analysis}

Topic T0 is the largest global topic (5{,}787 opinions)
and the strongest riser ($+1.34$ pp/decade). Its top
terms---\textit{sex offender}, \textit{rape},
\textit{sexual assault}, \textit{registration},
\textit{registry}---span both the underlying offenses
and the registration system built around them.
The evidence bundle included global topic opinions,
global-label topics and sub-theme topics, so the
LLM can recover what the statistical
models surfaced only structurally.

\begin{table}[t]
\centering
\small
\caption{T0 doctrinal arc: sex offense adjudication
and registration, 1970s--2020s.}
\label{tab:t0arc}
\setlength{\tabcolsep}{4pt}
\begin{tabular}{lp{1.5cm}p{3.4cm}}
\toprule
\textbf{Decades} & \textbf{Phase} & \textbf{Key developments} \\
\midrule
1970s--80s & Due process \& transfer &
  Older vocab: \textit{carnal knowledge}, \textit{sodomy};
  videotaped child interviews; no registration \\
1990s & Registration mandates &
  Wetterling Act; Megan's Law; community notification;
  SSODA, KSORA acronyms \\
2000s & AWA apparatus &
  Tier classification; 118-doc sub-theme topic;
  offense vs.\ registry vocabulary gap \\
2010s & Judicial pushback &
  Delinquency finding required; WSORA, SVORA;
  sealing petitions \\
2020s & GPS era &
  Satellite monitoring; lifetime satellite;
  hearsay scrutiny \\
\bottomrule
\end{tabular}
\end{table}

The LLM identifies a three-phase arc
(Table~\ref{tab:t0arc}). In the
\textbf{1970s and 1980s}, registration did not exist.
Courts focused on fair hearings and whether serious
cases should move to adult court. The sub-theme topics use older vocabulary---\textit{carnal knowledge},
\textit{sodomy}---that the global model no longer treats
as representative; the LLM confirms these cases are
about adjudication standards, not registration.
In the \textbf{1980s}, courts began accommodating
child victim testimony through videotaped interviews
and competency hearings---an era-specific procedural
vocabulary entirely absent from the global T0 label.

The \textbf{1990s mark the turning point}. Federal
laws required every state to create sex offender
registries (Wetterling Act; Megan's Law), and
courts were suddenly asked to apply adult-designed
rules to juveniles. The first registration-specific
sub-theme topics appear: community-notification
hearings, state statute acronyms (SSODA, KSORA).
Through the \textbf{2000s}, the Adam Walsh Act
the Adam Walsh Act imposed a national
tier-classification framework, producing the largest
sub-theme topic in the corpus---118 documents built
around registry and registration vocabulary while the
global T0 label centers on the offense.

In the \textbf{2010s}, courts pushed back:
some ruled registration could not be imposed without
a formal delinquency finding. In the \textbf{2020s},
two parallel streams emerge---GPS and lifetime
monitoring disputes, and stricter hearsay scrutiny in
sex offense trials. The LLM notes three major shifts:
confidential rehabilitation to public registration
(1970s to 1990s); flexible procedures to formal
evidentiary rules (1990s to 2010s); blanket
registration to individualized review (2010s onward).

\subsubsection{Topic T2: Juvenile Drug Trafficking}
\label{sec:t2analysis}

Topic T2 (2{,}212 opinions, $-0.25$ pp/decade) covers
the prosecution of juveniles for cocaine, heroin, crack
cocaine, and related narcotics. Despite its falling
prevalence trend, the LLM reveals that T2 is not one
story but four, driven by successive drug crises that
each produced new legal vocabulary and doctrine.

\begin{table}[t]
\centering
\small
\caption{T2 doctrinal arc: juvenile drug trafficking,
1970s--2020s.}
\label{tab:t2arc}
\setlength{\tabcolsep}{4pt}
\begin{tabular}{lp{1.5cm}p{3.4cm}}
\toprule
\textbf{Decades} & \textbf{Phase} & \textbf{Key developments} \\
\midrule
1970s & Rehabilitative &
  Parens patriae discretion; diversion to treatment;
  search \& seizure disputes \\
1980s--90s & War on drugs &
  Crack cocaine; mandatory minimums; school-zone
  enhancements; accomplice liability expanded \\
2000s & Rehab returns &
  Merger/allied-offenses doctrine; addiction
  considered in sentencing \\
2010s--20s & Opioid / fentanyl era &
  Heroin and fentanyl non-merge rule; racial
  disparity vocabulary enters opinions \\
\bottomrule
\end{tabular}
\end{table}

The LLM traces a four-phase arc
(Table~\ref{tab:t2arc}). In the
\textbf{1970s}, courts still operated under a
rehabilitative model---\textit{parens patriae}
discretion, diversion to treatment---but this was
already eroding. The \textbf{1980s and 1990s} brought
the war on drugs: crack cocaine and heroin drove
transfers to adult court, mandatory minimums, and
school-zone enhancements, with accomplice liability
expanded to cover juveniles in distribution networks.
Courts developed the merger and allied-offenses
doctrine---whether possession and distribution charges
should count as one offense or two for sentencing.

In the \textbf{2000s} some rehabilitative
considerations re-entered, especially for first-time
offenders, but trafficking remained harshly punished.
The structural break comes in the \textbf{2010s--2020s}
with the opioid crisis: fentanyl replaced heroin as
the paradigm substance, and courts developed a new
rule---heroin and fentanyl found mixed together do not
merge for sentencing because each poses distinct and
identifiable harms. This is a clear vocabulary and
doctrinal shift: \textit{allied offenses},
\textit{fentanyl}, \textit{separate and identifiable
harms} are all 2010s--2020s terms not present in the
topic's earlier vocabulary. The LLM also notes that
the 2020s produced the first explicit judicial
engagement with racial bias in drug sentencing, with
courts generally rejecting such claims but the
vocabulary of racial disparity entering appellate
opinions for the first time.

\subsubsection{Topic T6: Child Welfare and
Parental Substance Abuse}
\label{sec:t6analysis}

Topic T6 (791 opinions, $+0.88$ pp/decade) covers
termination of parental rights (TPR) and child removal
due to parental substance abuse. It is the
second-fastest riser in the corpus, and the LLM
analysis shows why: the topic's legal standard
changed fundamentally when the substance at its center
shifted from alcohol to methamphetamine.

\begin{table}[t]
\centering
\small
\caption{T6 doctrinal arc: child welfare and parental
substance abuse, 1970s--2020s.}
\label{tab:t6arc}
\setlength{\tabcolsep}{4pt}
\begin{tabular}{lp{1.5cm}p{3.4cm}}
\toprule
\textbf{Decades} & \textbf{Phase} & \textbf{Key developments} \\
\midrule
1970s--80s & Parental rights protective &
  Alcohol/heroin; remote-harm bar; cautious
  removal standard \\
1990s--2000s & ASFA \& meth shift &
  ASFA timelines; methamphetamine replaces alcohol
  as paradigm substance; ``children cannot wait'' \\
2010s--20s & Meth as presumptive unfitness &
  Bond exception narrowed; hair/cord-blood testing;
  newborn TPR cases \\
\bottomrule
\end{tabular}
\end{table}

The LLM identifies a three-phase arc
(Table~\ref{tab:t6arc}). In the
\textbf{1970s}, courts strongly protected parental
rights: removal required clear and convincing evidence
of actual harm, and a ``remote possibility'' of danger
was not enough. The doctrine was cautious, and courts
sometimes reversed removals where harm was speculative.

The \textbf{1990s and 2000s} mark the turning point.
ASFA imposed strict timelines on
family reunification, shifting the legal question from
``is the parent unfit?'' to ``can this child wait?''
Simultaneously, methamphetamine replaced alcohol as
the paradigm substance: courts found meth addiction
qualitatively more persistent and dangerous than
alcohol, and began treating ongoing meth use as
effectively incompatible with safe parenting. The
key doctrinal phrase---\textit{children cannot be
forced to await the maturity of their parents}---dates
from this era.

By the \textbf{2010s and 2020s}, the doctrine is
settled in a new place: courts explicitly name and
narrow the ``permissive bond exception'' (the
principle that an unusually strong parent-child bond
might outweigh TPR even with addiction), and apply
it rarely. The 2020s add a technological vocabulary
shift: hair follicle testing, umbilical cord blood
screening, and newborn testing positive for
methamphetamine enter the opinions as specific
evidence types, generating new sub-disputes over
test accuracy and secondhand exposure claims. The
LLM notes the sample is weighted toward Iowa courts,
which may reflect a jurisdiction with particularly
active TPR litigation; the principles are broadly
consistent with U.S. trends but not fully
representative of all state variations.


\section{Discussion}
\label{sec:discussion}

\subsection{Two Separate Systems in One Courthouse}
\label{sec:twosystems}

One pattern that stands out in the data is that
juvenile-related courts are handling two quite
different kinds of cases, each growing or shrinking
for different reasons. The first type involves a
young person accused of a crime---the traditional
delinquency case covering offenses, trials, and
sentencing. The second type involves the state
stepping in to protect a child from their own
family---the child welfare case covering neglect,
parental fitness, and foster care placement.

These two types of cases are governed by different
laws, involve different people, and respond to
different events in the outside world. The
delinquency side shrank over the study period,
consistent with falling youth arrest rates from
the mid-1990s onward and a broader shift toward
keeping young people out of the formal court system
where possible \citep{puzzanchera2022,
nationalresearchcouncil2013}. The child welfare
side grew substantially, driven by federal rules
set by ASFA, the drug epidemic's
effect on families, and new state family court
procedures. 

For anyone building tools or doing
research on juvenile justice as a single category,
this split matters: the field is not moving in one
direction---it is moving in two opposite directions
at the same time. The child welfare system has not
merely grown---the LLM analysis of Topic T6 shows
it developed its own body of doctrine, centred on
substance abuse evidence, ASFA permanency timelines,
and parental fitness standards, that shares almost
no vocabulary or legal logic with the delinquency
side. Because the two systems respond
to different legislative and social drivers, an
LLM prompted on a topic without specifying which
system it belongs to may retrieve evidence from
both and blend signals that should be interpreted
separately.

\subsection{What the Trends Actually Measure}
\label{sec:whattrends}

The numbers in this paper measure what courts were
asked to decide---not what actually happened to young
people. When child welfare cases triple their share of
the corpus, it means more families ended up in court,
not necessarily that abuse itself tripled.

A drop in a topic's share does not mean a problem
went away; it may mean the decision moved somewhere
harder to appeal.
The clearest example is judicial transfer (T3,
$-0.72$ pp/decade): states passed laws making
transfer automatic, removing the judge's discretion
and with it the basis for appeal.
Topic T2 (drug trafficking, $-0.25$ pp/decade) shows
a related pattern: the overall count fell while the
remaining cases grew harder---fentanyl-era litigation
raising questions that cocaine-era doctrine never
addressed.

A sub-theme cluster that falls outside any global
label is not a model error.
It is a signal that those courts were litigating
a question the rest of the corpus does not share.
The right response is closer inspection, not deletion.

\subsection{The Binary Trap}
\label{sec:binarytrap}

A recurring structural problem in juvenile justice
is that practitioners are often left with only two
choices when charging a young person: pursue a
consequence designed for adult offenders that feels
disproportionate for a juvenile, or retreat from
the charge entirely \citep{feld2017, schaefer2016}.
The corpus shows both sides of this trap.

T0 (sex offense and registration) shows what happens
when practitioners choose the first option.
Each federal registration mandate required states
to register juvenile offenders under lifetime
public frameworks designed for adult predators.
The result is visible in the data: thirty years of
fragmented, unsettled appeals as courts worked out
what each new mandate meant for young people.
The litigation grew because practitioners applied
registration; it never settled because registration
was designed for adult offenders, not juveniles
\citep{vincent2012, lipsey2009, peterson-badali2015}.

T4 (death penalty) shows the other side.
The topic peaked in the 1980s and fell steeply
through the 1990s and 2000s --- well before
\textit{Roper v.\ Simmons} formally ended the
juvenile death penalty in 2005.
That early decline reflects prosecutors choosing
not to seek a consequence they regarded as
constitutionally and morally untenable for young
defendants, long before the Court required them
to stop.
The conduct did not disappear; the charge did.

Both patterns carry the same warning for any tool
trained on this corpus: a trend line alone does not
tell you whether the underlying conduct changed or
whether practitioners changed what they were
willing to charge.

\subsection{Guidelines for AI Tools in Juvenile Justice}
\label{sec:aiwarning}

Juvenile justice agencies and courts increasingly use 
risk- and needs-assessment tools to inform supervision, 
service planning, detention, placement, custody, and 
disposition decisions for youth 
\citep{vincent2012, peterson-badali2015, mcphee2023}.
The corpus analyzed in this paper spans fifty years
of appellate opinions across both the delinquency
and child welfare systems. That breadth surfaces
five specific risks that any AI-based decision
support tool trained on such material must address.
Each risk emerges directly from what the data shows;
each points toward a concrete guideline that a 
tool should incorporate.

\paragraph{1.~Do not treat past patterns as current law.}
The trend data shows that courts routinely upheld
mandatory life-without-parole for juveniles through
the 2000s---a standard explicitly reversed by
\textit{Graham}, \textit{Miller}, and
\textit{Montgomery} after 2010.
A tool trained on the full arc without temporal
stratification will learn those earlier opinions
as settled doctrine, and may present an overturned
standard as current law
\citep{nist2023, whitehouse2022}.
The risk is not using historical case law---it is
failing to distinguish what the law was, from what
it is now. A tool must therefore be
temporally calibrated: it must identify which legal
standards have been superseded and weight or flag
opinions accordingly.

\paragraph{2.~Handle vocabulary drift across decades.}
The decade-local analysis in
Section~\ref{sec:decadelocal} shows that legal
vocabulary shifted substantially across the study
period. Terms like \textit{carnal knowledge} and
\textit{forcible rape} that dominated sex-offense
opinions in the 1970s gave way to \textit{sexual
assault} and \textit{sex offender registration}.
Topic T2 shows the same pattern by substance:
the 1980s--1990s vocabulary of \textit{crack
cocaine}, \textit{heroin}, and \textit{mandatory
minimum} gives way in the 2010s--2020s to
\textit{fentanyl}, \textit{allied offenses}, and
\textit{separate and identifiable harms}---terms
that did not exist in the topic's earlier decades.
This exposes the risk that a tool calibrated on
one era's vocabulary may systematically undercount
cases from other eras, even when the underlying
legal question is the same. A tool
must therefore align its embeddings and retrieval
indices to the vocabulary of the era being queried,
not to the aggregate corpus.

\paragraph{3.~Incorporate fragmented
jurisdiction-specific litigation.}
The corpus shows that sex offender
registration---the fastest-rising area of the entire
corpus---generates large sub-theme clusters in every
decade since 1994, each built around state-specific
statute acronyms (\textit{SSODA}, \textit{KSORA},
\textit{WSORA}, \textit{SVORA}) or federal-mandate
vocabulary (\textit{tier classification},
\textit{satellite monitoring}, \textit{lifetime
satellite}) that no global topic label captures.
This exposes the risk that a tool indexing opinions
by global labels may silently miss a substantial
share of registration litigation,
the area that most directly determines what
obligations a young person will carry for life.
A tool must therefore index and retrieve
sub-theme and jurisdiction-specific opinion clusters
in addition to global labels.

\paragraph{4.~Do not read trend lines as conduct trends.}
As Sections~\ref{sec:whattrends} and
\ref{sec:binarytrap} show, a rising or falling
trend in the corpus measures what courts were
asked to decide --- not how often the underlying
conduct occurred.
T3 (judicial transfer) fell sharply not because
fewer serious offenses occurred, but because
states made transfer automatic, removing the
judicial hearings that generated appeals.
T0 rose because registration generated decades
of appeals, not because sexual offending increased.
T4 fell because prosecutors stopped seeking the
death penalty for juveniles long before the law
required them to, not because serious offending
disappeared.
A tool must treat trend lines as measures of
legal activity, not as measures of juvenile
behaviour, and must not use them alone to
draw conclusions about how common or serious
an offense was in any given period.

\paragraph{5.~Design for two distinct legal systems.}
Section~\ref{sec:twosystems} shows that delinquency
and child welfare litigation have been moving in
opposite directions for decades and have developed
distinct bodies of doctrine with almost no shared
vocabulary or legal logic. This exposes the risk
that a tool trained on the combined corpus learns
a blended picture that accurately describes neither
system: for a sentencing question it may surface
child welfare cases, and for a parental rights
question it may surface offense cases.
A tool must therefore treat delinquency
and child welfare as separate legal domains, with
system-specific training, retrieval, and reasoning.


\section{Limitations and Future Work}
\label{sec:futurework}

Four limitations apply. First, 45.3\% of the 60,470
corpus opinions were assigned to the HDBSCAN outlier
cluster and excluded from all prevalence estimates; a
second-pass soft-assignment stage could recover a
meaningful fraction and test which trend findings are
sensitive to outlier handling. Second, cross-decade 
topic alignment uses a Jaccard threshold of 0.20 
over top-word sets, and the sub-theme rescue step 
recovers unmapped topics near that threshold. 
Since BERTopic produces topic embeddings directly, 
cosine similarity over those embeddings is an available 
alternative. Combining or replacing Jaccard alignment 
with embedding-based similarity \citep{reimers2019} 
would improve precision for topics whose vocabulary 
shifted across decades while the underlying doctrine 
remained stable. Comparing the two alignment strategies 
is a natural direction for future work. Third,
LLM outputs are informed hypotheses grounded in the
supplied opinions, not authoritative legal conclusions,
and periodic spot-checks against expert-coded samples
would provide a quality baseline.
Fourth, the 2020s decade covers only 2020--2025 and
all trend comparisons involving that period should
be treated as preliminary.

Beyond these fixes, three extensions offer the most
value. Supplementing CourtListener \citep{courtlistener}
with direct bulk downloads from state court administrative
offices or a commercial database such as Westlaw would
improve coverage of the 1970s and 1980s, where
older opinions are most underrepresented
\citep{livermore2017}. Weighting
opinions by citation count would let the pipeline
distinguish landmark decisions from routine ones.
Finally, applying the pipeline to a second legal
domain---family law, immigration courts, or
environmental enforcement---would test whether the
global--local modeling architecture generalizes
beyond juvenile justice.

\section{Conclusion}

The pipeline introduced here addresses the core challenge
of cross-decade legal corpus analysis: that topic models
fit on different time periods produce incompatible topic
inventories. Building a shared global topic space first,
then aligning decade-specific models via Jaccard
similarity, makes every mapping transparent and
checkable---and surfaces sub-theme topics that a single
global label cannot represent.

Across fifty years of U.S.\ juvenile justice appellate
case law, four patterns stand out: dramatic growth in
child welfare and sex offense registration litigation;
sharp decline in traditional punitive mechanisms; and a
rising sentencing-reform cluster driven by the
\textit{Graham}--\textit{Miller}--\textit{Montgomery}
constitutional transformation. These patterns are
invisible to close reading at this scale but fully
measurable, reproducible, and auditable with the approach
introduced here.

\section*{Ethics Statement}

All data is publicly available appellate opinions from
CourtListener. No private, sealed, or personally
identifying information about juveniles was accessed.
LLM outputs are framed as informed hypotheses grounded
in supplied evidence, not authoritative legal conclusions.
The pipeline is designed to support, not replace, human
legal review and produces no predictive claims about
individuals.

\bibliography{references}
\clearpage
\onecolumn
\appendix

\section{Major Constitutional and Statutory Inflection Points}
\label{app:inflections}
\input{appendix_a_inflection_points}

\section{Topic Modeling Model Comparison}
\label{app:modelcomp}
\input{appendix_b_model_comparison}

\section{Legal Theme Taxonomy}
\label{app:taxonomy}
\input{appendix_c_taxonomy}

\section{LLM Per-Topic Analysis: Full Outputs}
\label{app:pertopicoutputs}
\input{appendix_d_pertopic_outputs}

\end{document}

%% file: appendix_a_inflection_points.tex

This appendix summarises the key constitutional decisions and
federal statutes referenced throughout the paper.
Each entry identifies the year, the legal source, and its
doctrinal significance for the trends measured in the corpus.
Entries are listed chronologically.

\subsection*{Constitutional Decisions}

\paragraph{\textit{Kent v.\ United States}, 383 U.S.\ 541 (1966).}
The Supreme Court's first major juvenile justice ruling.
Held that before a juvenile court could waive jurisdiction
and transfer a young person to adult criminal court, it must
hold a hearing, provide access to counsel, and issue a
written statement of reasons. Established that juveniles
have due process rights in the transfer process. Its effect
is visible in the corpus as the Procedural and Evidentiary
theme begins at 12.6\% in the 1970s---elevated because
courts were still working through the implications of this
ruling and its immediate successor.

\paragraph{\textit{In re Gault}, 387 U.S.\ 1 (1967).}
Extended full procedural due process rights to juvenile
adjudications. The Court held that juveniles facing
delinquency proceedings must be given notice of charges,
the right to counsel, the right to confront witnesses,
and the privilege against self-incrimination. Together
with \textit{Kent}, \textit{Gault} transformed the juvenile
court from an informal, rehabilitative institution into one
bound by constitutional procedural requirements. Both
decisions predate the study window but directly shaped
the 1970s corpus.

\paragraph{\textit{In re Winship}, 397 U.S.\ 358 (1970).}
Held that the Due Process Clause requires proof beyond
a reasonable doubt for every fact necessary to constitute
a crime, and extended this standard to juvenile
delinquency proceedings. Before \textit{Winship},
many juvenile courts applied a lower preponderance-of-evidence
standard. The ruling established that juveniles facing
loss of liberty are entitled to the same evidentiary
protection as adults in criminal proceedings, completing
the due-process trilogy begun by \textit{Kent} and
\textit{Gault}.

\paragraph{\textit{Roper v.\ Simmons}, 543 U.S.\ 551 (2005).}
Held that the Eighth Amendment prohibits the execution of
individuals who committed their offenses before the age of
eighteen. Effectively ended the juvenile death penalty in
the United States. The sharp decline in Topic T4 (death
penalty, $-0.45$ pp/decade) tracks directly from this
ruling, which eliminated an entire category of appellate
litigation.

\paragraph{\textit{Graham v.\ Florida}, 560 U.S.\ 48 (2010).}
Held that sentencing a juvenile to life without the
possibility of parole for a non-homicide offense violates
the Eighth Amendment's prohibition on cruel and unusual
punishment. The first in the trilogy of rulings that
constitutionally constrained juvenile life sentences.

\paragraph{\textit{Miller v.\ Alabama}, 567 U.S.\ 460 (2012).}
Extended \textit{Graham} to hold that mandatory life without
parole is unconstitutional for all juvenile offenders,
regardless of the offense. Courts are required to consider
the individual's age, developmental history, and
circumstances before imposing the harshest sentence.
This ruling generated the sharp upward trend in
sentencing-reform cases visible after 2010 in the corpus.

\paragraph{\textit{Montgomery v.\ Louisiana}, 577 U.S.\ 190 (2016).}
Held that \textit{Miller} announced a new substantive rule
of constitutional law that applies retroactively, requiring
states to review sentences already imposed under the
mandatory LWOP standard. This extended the litigation wave
begun by \textit{Miller} by requiring courts to re-examine
thousands of existing sentences, sustaining the growth of
the sentencing-reform cluster through the 2020s.

\subsection*{Federal Statutes}

\paragraph{Juvenile Justice and Delinquency Prevention Act
(JJDPA), Pub.\ L.\ No.\ 93-415 (1974).}
Established the federal framework for juvenile justice
reform, requiring states receiving federal funding to
deinstitutionalise status offenders and remove juveniles
from adult jails. Shaped the decline of school discipline
and status offense litigation by redirecting those cases
away from formal court proceedings.

\paragraph{Child Abuse Prevention and Treatment Act
(CAPTA) Amendments of 1984, Pub.\ L.\ No.\ 98-457.}
Extended and strengthened the original 1974 CAPTA
framework, expanding mandatory child abuse reporting
obligations, tightening evidentiary standards for
child testimony, and increasing federal funding for
state child protective services. The 1984 amendments
coincided with a significant shift in how courts
handled child sexual abuse cases, including new
standards for the admissibility of child witness
testimony and the development of hearsay exceptions
for abuse victims. This drove modest growth in T0
(sex offense adjudication) during the 1980s---the
increase that precedes the sharp rise produced by
the sex offender registration mandates of the 1990s.

\paragraph{Jacob Wetterling Crimes Against Children and
Sexually Violent Offender Registration Act,
Pub.\ L.\ No.\ 103-322 (1994).}
Required states to establish sex offender registries.
The first of three successive federal mandates that each
produced a new wave of appellate litigation over
application to juvenile offenders, generating the
registration-specific sub-theme clusters documented in
Section~\ref{sec:decadelocal}.

\paragraph{Megan's Law, Pub.\ L.\ No.\ 104-145 (1996).}
Required public notification of sex offender registry
information. Introduced community-notification vocabulary
into the appellate record (\textit{SSODA}, \textit{KSORA})
that the global T0 label cannot capture, producing the
first large wave of sub-theme topics in the 1990s.

\paragraph{Adoption and Safe Families Act (ASFA),
Pub.\ L.\ No.\ 105-89 (1997).}
Imposed strict timelines on family reunification before
states could initiate proceedings to permanently terminate
parental rights. The single most consequential statutory
driver of the Child Welfare and Dependency theme's growth:
the theme's share jumps from 2.3\% to 9.5\% in the
decade following ASFA's enactment.

\paragraph{Adam Walsh Child Protection and Safety Act,
Pub.\ L.\ No.\ 109-248 (2006).}
Established a national, three-tier sex offender
classification system and expanded registration
requirements. Its enactment produced the largest
sub-theme cluster in the corpus: 118 documents from the
2000s decade built entirely around tier-classification
and registry management vocabulary that diverges from
the global T0 offense label.

%% file: appendix_b_model_comparison.tex

\subsection{Model Comparison Details}
\label{app:modelcomparison}

This appendix describes each topic modeling configuration
evaluated in the pilot comparison reported in
Section~\ref{sec:modeling}, followed by the results
that led to the selection of BERTopic with MPNet
embeddings.

\subsubsection{Word-Count-Based Models}

Word-count-based topic models work with frequency
representations of documents and are attractive as
starting points because they are fast, transparent,
and straightforward to aggregate by decade. Their
main limitation for legal text is that legal concepts
are often expressed through different wording across
courts and decades---the same doctrine can appear
under different statutory labels, in different
procedural postures, and with different terminology
in different jurisdictions. These models tend to
fragment a single legal concept into multiple topics
when surface word overlap is limited. Two baselines
were tested on the 6,000-document pilot subset.

\paragraph{Latent Dirichlet Allocation (LDA).}
LDA models each document as a mixture of latent
topics and each topic as a probability distribution
over the vocabulary \citep{blei2003}. It was treated
as a high-coverage baseline: it reliably produces
topic-word distributions and can be tracked over
time using standard mixture-weight aggregation. Its
main limitation was semantic fragmentation: the same
legal issue expressed with different wording across
courts and decades was often split into separate
topics. LDA achieved lower average coherence than
the embedding-based alternatives in the decade-slice
evaluations (mean c$_v$ = 0.382 $\pm$ 0.004).

\paragraph{Top2Vec with Doc2Vec embeddings.}
Top2Vec can be configured with Doc2Vec-style
embeddings to discover topic clusters without
requiring a fixed number of topics in advance
\citep{angelov2020}. This version surfaced
fine-grained subtopics in some configurations but
topic granularity and topic counts were sensitive
to hyperparameters and to corpus variation across
decades, making cross-decade comparability harder
to maintain. It was treated as an exploratory
baseline rather than a primary candidate.

\subsubsection{Embedding-Based Models}

Embedding-based models cluster documents using
transformer representations that encode meaning
beyond surface word overlap. This is better suited
to judicial opinions where boilerplate procedure
language is repeated across many cases while the
legally meaningful content can be expressed in
varied ways. Two candidates were evaluated.

\paragraph{BERTopic.}
BERTopic defines topics as clusters in a
document-embedding space and derives topic labels
using class-based TF-IDF (c-TF-IDF) computed over
the clustered texts \citep{grootendorst2022}.
Documents are embedded using a sentence-transformer
model \citep{reimers2019} to obtain dense semantic
representations. Embeddings are reduced with UMAP
\citep{mcinnes2018umap} and then clustered with
HDBSCAN \citep{mcinnes2017hdbscan}, which can also
mark outlier documents that do not belong to any
stable cluster. Topic descriptors are computed using
a vocabulary and c-TF-IDF weighting
\citep{pedregosa2011}. In this study, MPNet
embeddings were applied to the preprocessed,
phrase-aware corpus to produce topic signatures
that are both interpretable and easy to align across
decades.

\paragraph{Top2Vec with transformer embeddings.}
Top2Vec can also be paired with transformer
embeddings (e.g., all-MiniLM-L6-v2) to improve
semantic grouping \citep{angelov2020}. In the pilot
runs it produced coherent clusters in many slices,
but provided fewer built-in controls for
hierarchical reduction, standardized topic
identifiers, and export artifacts needed for the
multi-stage pipeline. It was treated as a contextual
comparator rather than as the primary modeling
backbone.

\subsubsection{Quantitative Results}

Model selection used three criteria: how coherent
and interpretable the topic descriptors were; how
distinct the topics were from one another; and how
well each model fit the pipeline's requirements for
hierarchical reduction, decade-local modeling, topic
alignment, and evidence export.

BERTopic with MPNet embeddings achieved the strongest
and most stable coherence across the 1980s--2020s
slices (mean c\_v = 0.493, SD 0.013, range
0.474--0.504 \citep{roder2015}), outperforming LDA
(0.382 $\pm$ 0.004), Top2Vec/MiniLM (0.467 $\pm$
0.035), and a LegalBERT variant (0.441 $\pm$ 0.058).
BERTopic/MPNet also showed the lowest topic
redundancy: mean nearest-neighbor cosine similarity
of approximately 0.40, compared to 0.66 for
Top2Vec/MiniLM, indicating more distinct topic
descriptors. Cross-decade continuity, measured by
best-match cosine similarity between topic word
signatures in adjacent decades, was strongest for
BERTopic/MPNet (mean $\approx$ 0.51; median
$\approx$ 0.58), exceeding Top2Vec/MiniLM
(mean $\approx$ 0.45) and Top2Vec/Doc2Vec
(mean $\approx$ 0.17). LDA showed high
adjacent-decade similarity (mean $\approx$ 0.60)
but its lower coherence and broader topic descriptors
reduced its usefulness for evidence-grounded legal
interpretation.

Operationally, BERTopic also supported the
pipeline's requirements: hierarchical topic
reduction for controlling granularity, reliable
export of document-topic assignments and c-TF-IDF
signatures for trend tables and topic alignment, and
straightforward assembly of evidence bundles for the
LLM-assisted stages. On the basis of these results,
BERTopic with MPNet embeddings was selected as the
primary modeling framework.

%% file: appendix_c_taxonomy.tex

Table~\ref{tab:fulltaxonomy} lists the complete 10-theme
taxonomy produced by the LLM classification described in
Section~\ref{sec:themeclassification}. Themes are ordered
by descending corpus coverage. Coverage is the share of
the 33,077 modeled opinions whose highest-probability
topic belongs to that theme. Trend is based on the
direction of the linear slope of the theme's
decade-normalized prevalence share. Top-5 topics are
listed by document count with the first two c-TF-IDF
representative terms.

\begingroup
\small
\setlength{\tabcolsep}{4pt}
\begin{longtable}{p{2.6cm} r r r p{5.5cm}}
\caption{Complete LLM-derived legal theme taxonomy.
Coverage = share of 33,077 modeled opinions.
Trend = direction of linear slope in decade-normalized
share. Topics = number of global topics assigned to
theme. Top-5 topics listed by document count with
representative c-TF-IDF terms.}
\label{tab:fulltaxonomy} \\
\toprule
\textbf{Theme} &
\textbf{Cov.} &
\textbf{Topics} &
\textbf{Trend} &
\textbf{Top-5 topics (ID: terms, docs)} \\
\midrule
\endfirsthead
\toprule
\textbf{Theme} &
\textbf{Cov.} &
\textbf{Topics} &
\textbf{Trend} &
\textbf{Top-5 topics (ID: terms, docs)} \\
\midrule
\endhead
\midrule
\multicolumn{5}{r}{\small\textit{Continued on next page}}\\
\endfoot
\bottomrule
\endlastfoot

Juvenile Offenses \& Adjudication
  & 46.5\%
  & 65
  & Stable
  & T0 (sex offender, rape, 5{,}787);
    T2 (cocaine, heroin, 2{,}212);
    T4 (death penalty, aggravating, 886);
    T10 (robbery, bank, 430);
    T12 (resist arrest, obstruct, 394) \\[4pt]

Procedural \& Evidentiary Issues
  & 15.3\%
  & 37
  & Falling
  & T3 (transfer, jurisdiction, 1{,}424);
    T5 (habeas corpus, extradition, 828);
    T29 (contempt, sanction, 199);
    T30 (discrimination, EEOC, 192);
    T42 (salary, debt, 158) \\[4pt]

Probation, Supervision \& Revocation
  &  9.6\%
  &  8
  & Rising
  & T1 (probation, supervision, 2{,}368);
    T17 (parole eligibility, Eighth Amendment, 291);
    T46 (revocation, escape, 148);
    T51 (proportionate penalty, postconviction, 136);
    T83 (parole eligibility, release date, 83) \\[4pt]

Child Welfare \& Dependency
  &  9.1\%
  & 17
  & Rising
  & T6 (methamphetamine, parental right, 791);
    T7 (neglect, adjudicator, 676);
    T13 (reunification, detrimental, 365);
    T26 (injurious environment, sexual abuse, 218);
    T41 (reunification, parental right affirmed, 159) \\[4pt]

Detention, Custody \& Placement
  &  5.5\%
  & 12
  & Rising
  & T8 (custody, caseworker, 640);
    T20 (inmate, cell, 251);
    T24 (secure detention, abscond, 232);
    T25 (permanency planning, nonsecure, 224);
    T45 (children services, orphan, 150) \\[4pt]

School Discipline, Truancy \& Status Offenses
  &  4.5\%
  & 16
  & Falling
  & T19 (unruly, school, 260);
    T33 (PINS, family court, 178);
    T36 (knife, locker search, 167);
    T56 (truancy, habitual, 127);
    T58 (student, locker search, 125) \\[4pt]

Mental Health \& Competency
  &  3.2\%
  &  8
  & Stable
  & T11 (sexually violent predator, behavioral, 413);
    T32 (hospitalization, involuntary commitment, 184);
    T35 (developmental disability, interpreter, 168);
    T79 (mental retardation, involuntary, 86);
    T84 (sanity commission, lineup, 81) \\[4pt]

Restitution \& Financial Sanctions
  &  2.6\%
  &  4
  & Stable
  & T9 (restitution, economic loss, 588);
    T47 (disciplinary, suspension, 145);
    T96 (nolo contendere, compensation fund, 67);
    T111 (moral turpitude, impeachable, 57) \\[4pt]

Traffic \& Motor Vehicle Offenses
  &  2.2\%
  &  7
  & Stable
  & T18 (drunk driving, license suspension, 266);
    T34 (vehicular homicide, blood alcohol, 172);
    T38 (driver license, motor vehicle, 165);
    T142 (negligent entrustment, collision, 36);
    T143 (governmental immunity, political subdivision, 36) \\[4pt]

Collateral \& Civil Consequences
  &  1.5\%
  &  8
  & Stable
  & T28 (alien, aggravated felony, 200);
    T93 (Indian tribe, ICWA notice, 69);
    T112 (immigration, special immigrant, 57);
    T123 (youth services, funding, 45);
    T129 (Indian reservation, tribal, 42) \\

\end{longtable}
\endgroup

%% file: appendix_d_pertopic_outputs.tex

This appendix reproduces verbatim the LLM-generated
per-topic trend analysis outputs for Topics T0, T2,
and T6. Each output was produced by supplying the
model with an evidence bundle drawn from globally
global-label topics and sub-theme topics in the relevant
sub-theme, covering all decades in which the topic
has coverage. The model received no information
beyond what was in that bundle. Outputs follow the
fixed prompt schema described in
Section~\ref{sec:pertopic}.

\subsection*{T0 --- Juvenile Sex Offense Adjudication and Registration (Trend Mode)}
\label{app:t0}

\begin{mdframed}[style=codeblock]
\nolinenumbers
\begin{verbatim}
## Topic Overview

Juvenile Sex Offense Adjudication and Registration: Doctrinal
Expansion and Procedural Complexity

This topic cluster concerns the adjudication of juveniles for sex
offenses (including rape, sexual assault, sexual contact, and
related conduct), the evidentiary and procedural rules governing
such cases, and the evolution of sex offender registration
requirements for juveniles. The prevalence of this topic shows a
rising and shifting trend: while the 1970s opinions focus on due
process and transfer to adult court, later decades increasingly
address registration, evidentiary rules, and the collateral
consequences of adjudication. The constitutional provisions most
frequently implicated are the Due Process Clause (Fourteenth
Amendment), Confrontation Clause (Sixth Amendment), and, in some
cases, the Eighth Amendment. Statutory drivers include state sex
offense statutes, sex offender registration acts (SORAs), and
rules of evidence (e.g., hearsay, prior bad acts).

## Decade-by-Decade Analysis

### 1970s

Dominant Legal Questions:
- Due process protections for juveniles accused of sex offenses
  (e.g., right to counsel, confrontation, notice).
- Standards and procedures for transfer to adult court in serious
  cases (e.g., rape).
- Evidentiary sufficiency and the role of prior juvenile records
  in sentencing.

Doctrinal Framework:
- Heavy reliance on In re Gault (1967) and Kent v. United States
  (1966) for due process and transfer standards.
- Courts emphasize that "an adjudicatory hearing in a Juvenile
  Court is not required to conform to all the requirements of a
  criminal trial for adults in order to comply with due process
  of law" (In Re Gantt).
- Early opinions do not address registration; focus is on
  adjudication and transfer.

Shifts in Reasoning:
- Gradual move from informal, rehabilitative juvenile proceedings
  toward more formal, adversarial process, especially in serious
  sex offense cases.
- Early recognition that "the seriousness of the alleged offense
  to the community and whether the protection of the community
  requires waiver" is a key factor in transfer decisions
  (Sherfield v. State).

Quotations:
- "The appellant had his constitutional rights to confrontation
  and cross-examination and all his 'due process of law' and
  other constitutional rights fully accorded within the
  parameters of In re Gault (1967), 387 U.S. 1..." (In Re Gantt).
- "The standards for certification stated in § 1112 is whether
  such juvenile or child is 'capable of knowing right from wrong,
  and to be held accountable for his acts.'" (Sherfield v.
  State).

### 1980s

Dominant Legal Questions:
- Expansion of evidentiary rules in juvenile sex offense cases
  (e.g., use of child testimony, hearsay, prior bad acts).
- Admissibility of videotaped interviews and the reliability of
  child witnesses.
- Early emergence of sex offender registration statutes, but not
  yet a dominant issue.

Doctrinal Framework:
- Application of rules of evidence to accommodate child victims
  (e.g., competency hearings, use of videotape).
- Courts begin to address the tension between protecting child
  victims and the rights of juvenile respondents.

Shifts in Reasoning:
- Increasing willingness to admit child testimony and out-of-
  court statements, provided reliability is established.
- Early recognition of the need for special procedures in child
  sex offense cases.

Quotations:
- "The trial judge has wide discretion in determining the
  competency of witnesses under 12 years of age." (State v.
  Ohrberg).

### 1990s

Dominant Legal Questions:
- Proliferation of sex offender registration statutes (SORAs) and
  their application to juveniles.
- The scope and constitutionality of registration requirements
  for juveniles adjudicated delinquent for sex offenses.
- Evidentiary issues: use of prior bad acts, hearsay, and
  confrontation rights.

Doctrinal Framework:
- Courts increasingly interpret and apply SORAs to juveniles,
  often with little statutory guidance.
- Some courts begin to recognize the unique status of juveniles
  and carve out exceptions or procedural protections.

Shifts in Reasoning:
- Movement from treating juvenile sex offense adjudications as
  confidential and rehabilitative to imposing adult-like
  collateral consequences (registration).
- Growing judicial discomfort with the breadth of registration
  requirements for juveniles, leading to statutory and
  constitutional challenges.

Quotations:
- "The only reference to juveniles in section 2 is in subsection
  (A)(5). Section 2(A)(5) thus offers enhanced protection for
  juveniles in that only juveniles who have been adjudicated
  delinquent for specified offenses fit section 2’s definition of
  'sex offender.'" (People v. S.B.).

### 2000s

Dominant Legal Questions:
- Implementation and challenge of juvenile sex offender
  registration.
- Procedural safeguards in juvenile sex offense adjudications
  (e.g., confrontation, hearsay, effective assistance).
- Dispositional issues: commitment to youth services, treatment,
  and proportionality.

Doctrinal Framework:
- Courts apply SORAs to juveniles, but some states (e.g.,
  Illinois) begin to recognize statutory or constitutional
  limits.
- Emphasis on balancing public safety, rehabilitation, and the
  unique status of juveniles.

Shifts in Reasoning:
- Increasing judicial recognition of the harshness and potential
  injustice of lifetime registration for juveniles.
- Some courts carve out exceptions or require individualized
  findings before imposing registration.

Quotations:
- "The clear import of section 3–5 is to afford juveniles
  enhanced protection under the Registration Act." (People v.
  S.B.).

### 2010s

Dominant Legal Questions:
- Constitutionality and proportionality of juvenile sex offender
  registration.
- Procedural due process in registration and adjudication.
- Evidentiary issues: confrontation, hearsay, prior bad acts, and
  the use of child testimony.

Doctrinal Framework:
- Some courts hold that registration cannot be imposed absent an
  adjudication of delinquency (People v. S.B.).
- Others uphold registration but require procedural safeguards
  and opportunities for relief.

Shifts in Reasoning:
- Growing skepticism about the efficacy and fairness of juvenile
  registration.
- Emphasis on individualized assessment and the possibility of
  relief from registration.

Quotations:
- "The result is a nondelinquent juvenile having fewer rights
  than a juvenile who was actually adjudicated delinquent in that
  the former has no ability to petition the circuit court to have
  his sex offender registration terminated..." (People v. S.B.).

### 2020s

Dominant Legal Questions:
- Continued litigation over the scope and fairness of juvenile
  sex offender registration.
- Application of confrontation and hearsay rules in juvenile sex
  offense trials.
- Sentencing proportionality and the role of prior bad acts.

Doctrinal Framework:
- Courts continue to grapple with the balance between public
  safety and the rehabilitative goals of juvenile justice.
- Some courts (e.g., State of Louisiana v. Keith Brown, II)
  scrutinize the use of prior bad acts and hearsay in light of
  confrontation rights.

Shifts in Reasoning:
- Increasing judicial willingness to find error in the admission
  of prior bad acts and hearsay, but often finding such errors
  harmless.
- Persistent tension between punitive and rehabilitative
  approaches.

Quotations:
- "It cannot be said that the introduction of this evidence,
  which Brown was unable to question or put into perspective
  given that A.O. was not present at trial or otherwise able to
  be cross-examined, was harmless beyond a reasonable doubt."
  (State of Louisiana v. Keith Brown, II, dissent).

## Cross-Decade Changes

Most Significant Doctrinal Shifts:

1. From Rehabilitation to Registration:
   - In the 1970s, juvenile sex offense adjudications were
   largely confidential and focused on rehabilitation. By the
   1990s and 2000s, the enactment of SORAs (e.g., Megan’s Law,
   state analogues) led to the imposition of public
   registration requirements on juveniles, often with little
   distinction from adults.
   - Key drivers: Federal and state SORAs (1990s–2000s), e.g.,
   Jacob Wetterling Act (1994), Megan’s Law (1996), Adam Walsh
   Act (2006).

2. Procedural Formalization:
   - Early opinions emphasized informality and flexibility;
   later decades saw increasing formalization, with full
   application of confrontation, hearsay, and effective
   assistance doctrines.
   - Key drivers: In re Gault (1967), Kent v. United States
   (1966), and subsequent Supreme Court and state court
   decisions.

3. Judicial Pushback and Individualization:
   - By the 2010s, some courts began to push back against
   blanket registration, requiring individualized findings or
   providing mechanisms for relief (e.g., People v. S.B.,
   2011).
   - Key drivers: State legislative amendments, constitutional
   challenges (due process, equal protection), and empirical
   critiques of registration’s efficacy.

4. Evidentiary Evolution:
   - Increasing accommodation for child victim testimony
   (e.g., videotaped interviews, relaxed hearsay rules), but
   also heightened scrutiny of confrontation rights post-
   Crawford v. Washington (2004).

## Drivers of the Trend

Legal Drivers:
- Enactment and expansion of sex offender registration statutes,
  often in response to high-profile crimes and federal mandates.
- Supreme Court decisions clarifying juvenile due process rights
  (Gault, Kent, Winship), and later, confrontation rights
  (Crawford).
- State appellate decisions interpreting the scope and
  application of SORAs to juveniles, sometimes limiting their
  reach.

Institutional/Social Drivers:
- Public concern and political pressure regarding child sexual
  abuse and recidivism.
- Media coverage and legislative responses to notorious cases
  (e.g., Megan Kanka, Jacob Wetterling).
- Shifts in juvenile justice philosophy: from rehabilitation to
  public safety and back toward individualized assessment.

Opinion Evidence:
- Multiple opinions cite legislative changes and public policy
  concerns as motivating factors for expanding registration.
- Some opinions (e.g., People v. S.B.) explicitly note the
  tension between legislative intent to protect the public and
  the need for enhanced protections for juveniles.

## Decade-Local Specifics Not Fully Surfaced at Global Level

- 1970s: Focus on transfer to adult court and due process, with
  little mention of registration or collateral consequences.
- 1990s: Early judicial discomfort with applying adult
  registration schemes to juveniles, leading to some statutory
  carve-outs or procedural protections.
- 2000s–2010s: Some states (e.g., Illinois) create statutory
  mechanisms for juveniles to petition for removal from the
  registry, reflecting changing social norms and empirical
  research on juvenile recidivism.
- 2020s: Heightened scrutiny of evidentiary practices (e.g.,
  prior bad acts, hearsay) in light of confrontation rights, and
  continued debate over the proportionality of registration and
  sentencing.

These features reflect evolving social attitudes toward juvenile
culpability, the impact of neuroscience on perceptions of
adolescent development, and the increasing sophistication of both
prosecution and defense in litigating these cases.

## Key Cases and Their Role in the Evolution

1. In Re Gantt (1978, 1970s):
   - Early articulation of due process rights in juvenile sex
   offense adjudications.
   - "The appellant had his constitutional rights to
   confrontation and cross-examination and all his 'due
   process of law' and other constitutional rights fully
   accorded within the parameters of In re Gault (1967), 387
   U.S. 1..."

2. Sherfield v. State (1973, 1970s):
   - Application of Kent standards to transfer decisions in
   juvenile rape cases.
   - "The standards for certification stated in § 1112 is
   whether such juvenile or child is 'capable of knowing right
   from wrong, and to be held accountable for his acts.'"

3. People v. S.B. (2011, 2010s):
   - Illinois appellate court holds that registration cannot
   be imposed on a juvenile absent an adjudication of
   delinquency.
   - "The only reference to juveniles in section 2 is in
   subsection (A)(5). Section 2(A)(5) thus offers enhanced
   protection for juveniles..."

4. State v. Matha (1995, 1990s):
   - Upholds indefinite commitment for juvenile sex offenders,
   reflecting the punitive turn of the 1990s.
   - "The court determined, due to the serious nature of the
   crimes, Matha needed intensive counselling."

5. State of Louisiana v. Keith Brown, II (2024, 2020s):
   - Addresses the admissibility of prior bad acts and
   confrontation rights in juvenile sex offense trials.
   - "It cannot be said that the introduction of this
   evidence, which Brown was unable to question or put into
   perspective given that A.O. was not present at trial or
   otherwise able to be cross-examined, was harmless beyond a
   reasonable doubt." (dissent)

6. In re D.J. (2020, 2020s):
   - Upholds the use of prior acts as evidence of grooming in
   juvenile sex offense adjudications.
   - "The evidence of appellant's prior instances of watching
   a pornographic movie and engaging in a continuing course of
   sexual activity with the victim was relevant to and used
   for the legitimate purpose of showing appellant's
   opportunity, plan, and his preparation of the victim for
   later sexual activity, i.e. grooming of the victim."

## Boundary Cases and Evidence Limits

- People v. S.B. complicates the trend by holding that a juvenile
  found "not not guilty" (i.e., not acquitted, but not
  adjudicated delinquent) cannot be required to register,
  highlighting the importance of procedural posture.
- State of Louisiana v. Keith Brown, II (dissent) raises serious
  concerns about the use of hearsay and confrontation violations
  in admitting prior bad acts, suggesting that not all courts are
  comfortable with the broad admissibility of such evidence.
- The sample is limited in its coverage of federal constitutional
  challenges to registration, and does not include Supreme Court
  decisions such as Smith v. Doe (2003) or In re Gault itself,
  but the opinions do cite and apply these doctrines.

What the sample cannot tell us:
- The full range of state legislative responses to juvenile
  registration post-2010s.
- The impact of recent Supreme Court decisions on juvenile
  sentencing (e.g., Miller v. Alabama, Graham v. Florida) on sex
  offense registration.
- Empirical outcomes for juveniles subject to registration.

What the sample does show:
- A clear doctrinal arc from informal, rehabilitative proceedings
  to formal, punitive, and registration-driven consequences, with
  some recent judicial pushback and calls for individualized
  assessment and procedural fairness.
\end{verbatim}
\end{mdframed}

\subsection*{T2 --- Juvenile Drug Trafficking (Trend Mode)}
\label{app:t2}

\begin{mdframed}[style=codeblock]
\nolinenumbers
\begin{verbatim}
## Topic Overview

Juvenile Drug Distribution and Trafficking: Doctrinal Evolution
and Sentencing

This topic cluster concerns the prosecution, adjudication, and
sentencing of juveniles (and young adults) for offenses involving
the distribution, trafficking, or possession with intent to
distribute controlled substances—primarily cocaine, heroin, crack
cocaine, and related narcotics. The opinions span from the 1970s
through the 2020s and reflect a shifting but persistent
prevalence, with a notable rise in the 1980s and 1990s, a plateau
in the 2000s, and continued significance into the 2010s and
2020s. The cluster implicates constitutional provisions (notably
the Fourth, Fifth, and Fourteenth Amendments), state and federal
drug statutes (e.g., Controlled Substances Act, state analogues),
sentencing regimes, and doctrines concerning juvenile transfer,
accomplice liability, and the merger of offenses.

## Decade-by-Decade Analysis

### 1970s

Dominant Legal Questions:
- The appropriateness of rehabilitative versus punitive responses
  to juvenile narcotics offenses.
- The constitutionality of search and seizure practices in
  narcotics investigations.
- The scope of judicial discretion in diverting drug-addicted
  offenders to treatment rather than incarceration.

Doctrinal Framework:
- Courts often balanced the parens patriae philosophy of juvenile
  justice with the rising public concern over narcotics.
- Early cases (e.g., People v. Meza) upheld broad judicial
  discretion to deny diversion to treatment programs, emphasizing
  public safety and recidivism:
  > "Repeated violations of the narcotics laws are not
  irrelevant in the defendant’s pattern of criminality."
  (People v. Meza)

- Search and seizure doctrine was in flux, with courts wrestling
  with the exigency exception and the plain view doctrine in
  narcotics cases involving juveniles (Commonwealth v. Forde).

Visible Shifts:
- The decade saw the beginning of a shift from purely
  rehabilitative approaches to more punitive responses for repeat
  or serious juvenile drug offenders.

### 1980s

Dominant Legal Questions:
- The expansion of accomplice and conspiracy liability for
  juveniles in drug distribution networks.
- The use of informants and undercover operations in juvenile
  drug cases.
- The rise of mandatory minimums and sentencing enhancements for
  drug offenses.

Doctrinal Framework:
- Courts increasingly treated juveniles as adults in serious drug
  cases, especially where distribution or trafficking was
  charged.
- The reliability of informant testimony and the sufficiency of
  circumstantial evidence for accomplice liability were central:
  > "Presence at the scene of a crime is a fact which, together
  with other facts, may support a finding that the defendant
  acted as an accomplice." (State v. Pronovost)

Visible Shifts:
- The "war on drugs" era led to more frequent transfer of
  juveniles to adult court and harsher sentencing, with less
  emphasis on individualized treatment.

### 1990s

Dominant Legal Questions:
- The constitutionality and scope of juvenile transfer statutes
  for drug trafficking offenses.
- The merger of possession and distribution offenses for
  sentencing purposes.
- The use of prior juvenile adjudications in adult sentencing for
  drug crimes.

Doctrinal Framework:
- Courts applied the "same evidence" and "same conduct" tests to
  determine whether multiple drug charges should merge for
  sentencing (Commonwealth v. Wright).
- The use of prior juvenile adjudications as sentencing
  enhancements was upheld, provided statutory requirements were
  met (State v. Wade).

Visible Shifts:
- The decade saw the solidification of punitive approaches, with
  courts upholding lengthy sentences for juveniles convicted of
  drug distribution, especially where aggravating factors (e.g.,
  proximity to schools, use of firearms) were present.

### 2000s

Dominant Legal Questions:
- The proportionality of sentences for juvenile drug trafficking
  in light of evolving Eighth Amendment jurisprudence.
- The role of addiction and rehabilitation in sentencing.
- The application of allied offenses/merger doctrine to multiple
  drug charges arising from a single transaction.

Doctrinal Framework:
- Courts continued to uphold significant sentences for juvenile
  drug distribution, but some opinions reflected concern for
  rehabilitation and the impact of addiction (State v. Walley).
- The merger doctrine was applied inconsistently, with some
  courts merging possession and distribution charges, while
  others did not, depending on the facts and statutory language
  (State v. Daniels).

Visible Shifts:
- There was a gradual reintroduction of rehabilitative
  considerations, especially for first-time or low-level
  offenders, but serious trafficking offenses continued to draw
  harsh penalties.

### 2010s

Dominant Legal Questions:
- The impact of the opioid crisis on sentencing for juvenile
  heroin/fentanyl distribution.
- The use of "high crime area" evidence and its prejudicial
  effect in juvenile drug trials.
- The continued tension between rehabilitation and punishment in
  juvenile sentencing.

Doctrinal Framework:
- Courts scrutinized the admissibility of evidence regarding
  neighborhood crime rates, finding such evidence often unduly
  prejudicial (J.R. v. State).
- Sentencing courts balanced addiction, youth, and recidivism in
  determining whether to impose incarceration or alternative
  sentences (State v. Boykin, State v. Cathey).

Visible Shifts:
- The opioid crisis led to increased sentences for distribution
  of heroin/fentanyl, with courts emphasizing the public health
  impact:
  > "That amount of drugs put on the street could be a mass
  devastation." (State v. Daniels)

### 2020s

Dominant Legal Questions:
- The merger of trafficking offenses for different drugs found in
  a single mixture (e.g., heroin/fentanyl).
- The role of socioeconomic and racial factors in sentencing for
  drug distribution.
- The continued use of mandatory minimums and the application of
  allied offenses doctrine.

Doctrinal Framework:
- Courts have grappled with whether trafficking in heroin and
  trafficking in fentanyl (when mixed together) are allied
  offenses that must merge for sentencing. The majority view is
  that they do not merge, as each requires proof of a different
  substance and reflects distinct societal harms (State v.
  Daniels, State v. Howard).
- Some courts have addressed, and generally rejected, claims that
  sentencing was improperly influenced by race or socioeconomic
  status (State v. Carr).

Visible Shifts:
- The doctrine has become more nuanced, with some dissenting
  opinions urging merger of trafficking offenses for mixed drugs,
  but the prevailing trend is to treat each as a separate
  offense, especially in the context of the opioid epidemic.

## Cross-Decade Changes

1. Shift from Rehabilitation to Punishment (1970s–1990s):
- Early opinions emphasized rehabilitation and individualized
  treatment for juvenile drug offenders. By the 1980s and 1990s,
  the "war on drugs" and public concern over crack cocaine and
  heroin led to more punitive approaches, including frequent
  transfer to adult court and mandatory minimum sentences.

2. Expansion of Accomplice and Conspiracy Liability:
- Courts broadened the scope of accomplice and conspiracy
  liability for juveniles involved in drug distribution networks,
  often relying on circumstantial evidence and informant
  testimony.

3. Sentencing Enhancements and Merger Doctrine:
- The 1990s and 2000s saw the proliferation of sentencing
  enhancements (e.g., proximity to schools, use of firearms,
  prior convictions) and the development of the merger/allied
  offenses doctrine, with courts increasingly treating possession
  and distribution as separate offenses unless the facts
  compelled merger.

4. Opioid Crisis and Fentanyl (2010s–2020s):
- The rise of fentanyl and synthetic opioids led to even harsher
  sentencing for distribution offenses, with courts emphasizing
  the distinct and severe harms posed by each substance, often
  refusing to merge trafficking convictions for heroin and
  fentanyl even when found in a single mixture.

5. Racial and Socioeconomic Factors:
- Recent opinions have addressed, and generally rejected, claims
  that sentencing was improperly influenced by race or class,
  reaffirming the principle that such factors must not be
  considered.

## Drivers of the Trend

Legal Drivers:
- Statutory changes: The Controlled Substances Act (1970), state
  analogues, and the proliferation of mandatory minimums and
  sentencing enhancements.
- Supreme Court and appellate decisions clarifying the merger
  doctrine, accomplice liability, and the scope of juvenile
  transfer statutes.
- The opioid crisis prompted legislative and judicial responses
  emphasizing the distinct harms of fentanyl and heroin.

Institutional/Social Drivers:
- The "war on drugs" and public concern over crack cocaine and
  heroin in the 1980s–1990s led to more punitive approaches.
- The opioid epidemic in the 2010s–2020s drove courts to treat
  fentanyl and heroin as distinct for sentencing, reflecting
  public health concerns.
- Persistent racial disparities in drug enforcement and
  sentencing have prompted judicial scrutiny, though courts
  generally reject claims of explicit bias in individual cases.

## Decade-Local Specifics Not Fully Surfaced at Global Level

- 1970s: Courts still operated under the parens patriae model,
  with significant judicial discretion to divert juveniles to
  treatment, but this was eroding.
- 1980s: The use of informants and undercover operations became
  central, with courts developing doctrines for the reliability
  and sufficiency of such evidence.
- 1990s: The merger doctrine was in flux, with some courts
  merging possession and distribution, others not, depending on
  statutory language and facts.
- 2010s–2020s: The opioid crisis led to a focus on the unique
  dangers of fentanyl, with courts often refusing to merge
  trafficking convictions for heroin and fentanyl, even when
  mixed.

## Key Cases and Their Role in the Evolution

1. People v. Meza (1971, 1970s):
   - Upheld broad judicial discretion to deny diversion to
   treatment for repeat juvenile narcotics offenders:
     > "Repeated violations of the narcotics laws are not
     irrelevant in the defendant’s pattern of criminality."

2. Commonwealth v. Forde (1975, 1970s):
   - Addressed the exigency exception and plain view doctrine
   in narcotics searches involving juveniles:
     > "We hold here only that where the exigency is
     reasonably foreseeable and the police offer no
     justifiable excuse for their prior delay in obtaining a
     warrant, the exigency exception to the warrant
     requirement is not open to them."

3. State v. Pronovost (1984, 1980s):
   - Clarified accomplice liability for juveniles in drug
   distribution:
     > "Presence at the scene of a crime is a fact which,
     together with other facts, may support a finding that
     the defendant acted as an accomplice."

4. State v. Daniels (2020, 2020s):
   - Refused to merge trafficking in heroin and trafficking in
   fentanyl for sentencing, emphasizing distinct harms:
     > "We cannot overstate the harm that fentanyl has
     wrought on this state... trafficking in heroin and
     trafficking in fentanyl pose separate and identifiable
     harms under Ruff and do not merge as allied offenses."

5. State v. Howard (2023, 2020s):
   - Affirmed convictions for trafficking in multiple drugs,
   upholding constructive possession and the refusal to merge
   offenses:
     > "Collectively, this evidence established that Howard
     was conscious of the presence of the drugs and
     exercised dominion and control over them."

6. State v. Carr (2021, 2020s):
   - Addressed, and rejected, claims that sentencing was
   improperly influenced by race or socioeconomic status:
     > "The trial court, in fact, did the exact opposite by
     specifically stating that it had made a concerted
     effort to ignore Carr’s race and the concept of 'white
     privilege' when issuing its sentencing decision."

## Boundary Cases and Evidence Limits

- Boundary Cases:
  - State v. Daniels (2020) includes a strong dissent arguing
  that trafficking in heroin and fentanyl (when mixed) should
  merge for sentencing, reflecting a minority view that may
  gain traction as the law evolves.
  - Some 2010s cases (e.g., J.R. v. State) highlight the
  prejudicial effect of "high crime area" evidence, suggesting
  ongoing concerns about fairness in juvenile drug
  prosecutions.

- Evidence Limits:
  - The sample is limited to appellate opinions and may
  underrepresent cases resolved at the trial level or through
  diversion.
  - The opinions reflect the law as applied in particular
  jurisdictions and may not capture all regional or statutory
  variations.
  - The sample may understate the impact of recent Supreme
  Court decisions (e.g., Miller v. Alabama on juvenile
  sentencing) if not directly cited in the opinions.

---

In summary:
Across five decades, the doctrine governing juvenile drug
distribution and trafficking has evolved from a rehabilitative,
discretionary model to a more punitive, structured regime, with
increasing emphasis on public safety, sentencing enhancements,
and the distinct harms of specific narcotics (notably fentanyl).
The merger doctrine has become more restrictive, especially in
the context of the opioid crisis, and courts have generally
rejected claims that sentencing is improperly influenced by race
or class. The trend is driven by statutory changes, social
phenomena (the "war on drugs," opioid epidemic), and evolving
judicial doctrine, with some boundary cases suggesting possible
future shifts.
\end{verbatim}
\end{mdframed}

\subsection*{T6 --- Child Welfare and Parental Substance Abuse (Trend Mode)}
\label{app:t6}

\begin{mdframed}[style=codeblock]
\nolinenumbers
\begin{verbatim}
## Topic Overview

Termination of Parental Rights for Methamphetamine Use

This topic cluster centers on the legal doctrine and practice of
terminating parental rights (TPR) or removing children from
parental custody due to parental methamphetamine use, substance
abuse, and related failures to maintain sobriety, provide
nurturing care, or engage with court-ordered assistance. The
trend is rising in prevalence, especially from the 2000s onward.
The constitutional and statutory provisions most frequently at
issue are the Due Process Clause (14th Amendment), state child
welfare statutes (e.g., Iowa Code § 232.116), and the "best
interests of the child" standard. The doctrine involves balancing
parental rights against the state's parens patriae interest in
child welfare, with increasing emphasis on clear and convincing
evidence of harm or risk due to parental substance abuse.

## Decade-by-Decade Analysis

### 1970s

Dominant Legal Questions:
- What evidentiary standard is required to remove children from
  parents due to drug use?
- Does mere presence of drugs or parental substance use justify
  removal/TPR?

Doctrinal Framework:
- Strong protection of parental rights; removal requires clear
  and convincing evidence of actual harm or imminent risk.
- Courts distinguish between "remote possibility" of harm and
  actual risk.

Shifts in Reasoning:
- Courts are reluctant to terminate rights absent concrete
  evidence of harm:
  > "There is, as the court below found, a 'remote possibility'
  that the children may be endangered by their present
  environment but remote possibilities do not provide grounds
  sufficient for removing a child from parental custody." (In
  re WO, 1979)
- Dissenting voices begin to frame parental drug use as a form of
  child abuse.

### 1980s

Dominant Legal Questions:
- What constitutes unfitness due to substance abuse?
- How much time/opportunity must be given for parental
  rehabilitation?

Doctrinal Framework:
- Statutory frameworks (e.g., Iowa Code § 232.116) require clear
  and convincing evidence that the child cannot be safely
  returned.
- Emphasis on the best interests of the child and the need for
  permanency.

Shifts in Reasoning:
- Courts increasingly recognize substance abuse as a barrier to
  reunification:
  > "Her abuse of alcohol has proven to be the principal
  impediment to establishing a stable home for her children."
  (E.J. v. State, 1989)
- The passage of time and repeated failures at sobriety weigh
  heavily against parents.

### 1990s

Dominant Legal Questions:
- How do courts balance parental progress in treatment against
  the child's need for stability?
- What is the role of state-provided services and reasonable
  efforts?

Doctrinal Framework:
- Continued reliance on clear and convincing evidence.
- Focus on whether the parent can provide a safe home "at the
  present time."
- Reasonable efforts by the state are required, but parents must
  also engage.

Shifts in Reasoning:
- Courts become less tolerant of repeated relapses and more
  willing to terminate rights after failed reunification
  attempts:
  > "Children cannot be forced to await the maturity of their
  parents." (In the Interest of S.A., 1993)
- Substance abuse, especially methamphetamine, is increasingly
  seen as incompatible with safe parenting.

### 2000s

Dominant Legal Questions:
- What is the threshold for "reasonable efforts" by the state?
- How do courts handle cases where parents make some progress but
  relapse?

Doctrinal Framework:
- Statutory grounds for TPR are strictly applied.
- The best interests of the child and permanency are paramount.

Shifts in Reasoning:
- Courts emphasize that "time is a critical element" and that
  children should not wait indefinitely for parental
  rehabilitation:
  > "We cannot deprive a child of permanency after the State
  has proved a ground for termination... by hoping someday a
  parent will learn to be a parent." (In re P.L., 2009)
- Methamphetamine use is treated as particularly dangerous and
  persistent.

### 2010s

Dominant Legal Questions:
- How do courts weigh the parent-child bond against ongoing
  substance abuse?
- What exceptions, if any, allow for continued parental rights
  despite addiction?

Doctrinal Framework:
- Continued focus on clear and convincing evidence, best
  interests, and statutory exceptions (e.g., closeness of bond).
- Courts require evidence that the bond outweighs the risk and
  need for permanency.

Shifts in Reasoning:
- Courts are explicit that love or a bond is not enough to
  prevent termination if substance abuse persists:
  > "Neither a parent’s love nor the mere existence of a bond
  is enough to prevent termination." (In re D.W., 2010)
- Methamphetamine and opioid epidemics drive more frequent
  findings of unfitness.

### 2020s

Dominant Legal Questions:
- What is the impact of repeated failed treatment and relapses on
  TPR?
- How do courts handle cases involving newborns testing positive
  for methamphetamine?

Doctrinal Framework:
- Statutory grounds for TPR are strictly enforced; courts require
  evidence that the parent cannot provide a safe home now or in
  the foreseeable future.
- The "permissive bond exception" is rarely applied unless the
  bond is exceptionally strong and the parent is otherwise fit.

Shifts in Reasoning:
- Courts are increasingly unwilling to grant extensions or
  additional time for reunification where methamphetamine use is
  ongoing:
  > "The mother made virtually no progress over the life of the
  case and there is no reason to think that the need for
  removal... would be resolved within six months." (In the
  Interest of T.M.-L., 2025)
- Methamphetamine use during pregnancy and at birth is treated as
  strong evidence of unfitness.

## Cross-Decade Changes

Significant Doctrinal Shifts:
- 1970s–1980s: Shift from requiring actual harm to accepting
  imminent risk from substance abuse as sufficient for
  removal/TPR (see In re WO, 1979 vs. E.J. v. State, 1989).
- 1990s–2000s: Increasingly strict application of statutory
  timelines and best interests analysis; less tolerance for
  repeated relapses (see In the Interest of S.A., 1993; In re
  P.L., 2009).
- 2010s–2020s: Methamphetamine use, especially during pregnancy
  or with newborns, is treated as presumptively incompatible with
  safe parenting. The "permissive bond exception" is rarely
  applied unless the parent is demonstrably sober and stable.

Drivers of Change:
- State and federal legislation (e.g., Adoption and Safe Families
  Act, 1997) emphasizing permanency and timelines.
- Growing recognition of the dangers of methamphetamine and
  opioid use.
- Case law such as In re A.B., 815 N.W.2d 764 (Iowa 2012),
  affirming that unresolved, severe, and chronic drug addiction
  can render a parent unfit.

## Drivers of the Trend

Legal Drivers:
- Statutory changes requiring timely permanency decisions (e.g.,
  ASFA).
- Case law clarifying that substance abuse, especially
  methamphetamine, is a sufficient basis for TPR if it endangers
  the child or prevents safe parenting.
- Increasingly strict interpretation of "reasonable efforts" and
  "best interests" standards.

Institutional/Social Drivers:
- Methamphetamine and opioid epidemics leading to more cases
  involving parental substance abuse.
- Greater awareness of the impact of prenatal drug exposure and
  early childhood trauma.
- Policy shifts toward child safety and permanency over parental
  rehabilitation.

## Decade-Local Specifics Not Fully Surfaced at Global Level

- 1970s: Courts still sometimes view substance use as a medical
  issue rather than a per se ground for unfitness (see In re WO,
  1979).
- 1990s: Greater attention to the adequacy of state-provided
  services and the procedural rights of parents.
- 2020s: Increased use of hair, patch, and umbilical cord drug
  testing at birth; more cases involving newborns testing
  positive for methamphetamine; more explicit judicial skepticism
  of parents' explanations for positive tests.

## Key Cases and Their Role in the Evolution

1. In re WO (Cal. Ct. App. 1979, 1970s):
   - Early articulation of the "clear and convincing evidence"
   standard and skepticism of removal based on "remote
   possibility" of harm.
   - "Remote possibilities do not provide grounds sufficient
   for removing a child from parental custody."

2. E.J. v. State (Iowa 1989, 1980s):
   - Substance abuse as a principal impediment to
   reunification; clear and convincing evidence required for
   TPR.
   - "Her abuse of alcohol has proven to be the principal
   impediment to establishing a stable home for her children."

3. In the Interest of S.A. (Iowa 1993, 1990s):
   - Emphasizes that children cannot be forced to wait for
   parental maturity.
   - "Children cannot be forced to await the maturity of their
   parents."

4. In re P.L. (Iowa 2009, 2000s):
   - Explicit statement that courts cannot deprive a child of
   permanency by hoping for future parental rehabilitation.
   - "We cannot deprive a child of permanency after the State
   has proved a ground for termination... by hoping someday a
   parent will learn to be a parent."

5. In the Interest of T.M.-L. (Iowa Ct. App. 2025, 2020s):
   - Methamphetamine use at birth, repeated failed treatment,
   and lack of progress as grounds for TPR.
   - "Every single test result has been positive for
   methamphetamine... The mother made virtually no progress
   over the life of the case and there is no reason to think
   that the need for removal... would be resolved within six
   months."

6. In the Interest of R.D. (Iowa Ct. App. 2024, 2020s):
   - Both parents with long-term substance abuse histories;
   TPR affirmed despite requests for more time and relative
   placement.
   - "Considering the mother’s substance-use history, failed
   past attempts at sobriety, and lack of engagement in
   services during these proceedings, we cannot find the need
   for removal would be eliminated if the mother was given
   more time."

## Boundary Cases and Evidence Limits

- In re WO (1979): The majority and dissent illustrate the
  tension between parental rights and child safety, with the
  majority requiring actual harm and the dissent arguing that
  exposure to drug use is itself a form of abuse.
- In the Interest of S.A. (1993): The court acknowledges the
  difficulty of TPR cases and the lack of guarantees that
  children will find satisfactory adoptive homes.
- Recent 2020s cases: Some parents contest the accuracy of drug
  tests or claim secondhand exposure, but courts are increasingly
  skeptical of such defenses.
- Sample limits: The opinions are heavily weighted toward Iowa
  and similar Midwestern states, which may reflect local
  statutory frameworks and social conditions. The trend may be
  less pronounced in jurisdictions with different statutory or
  cultural approaches.

What this sample cannot tell us:
- The full range of outcomes in cases where parents successfully
  rehabilitate.
- The impact of new treatment modalities or family preservation
  programs not reflected in appellate opinions.
- How courts handle cases involving substances other than
  methamphetamine, or where substance abuse is not the primary
  issue.

---

In summary:
Over the past five decades, U.S. juvenile courts have moved from
a cautious, rights-protective approach to parental substance
abuse toward a doctrine that treats methamphetamine
addiction—especially when persistent or present at birth—as a
strong, often sufficient ground for termination of parental
rights. Statutory changes, the methamphetamine epidemic, and
evolving judicial attitudes have driven a marked increase in the
prevalence and doctrinal clarity of this topic.
\end{verbatim}
\end{mdframed}